\documentclass[12pt, draftclsnofoot, onecolumn]{IEEEtran}
\usepackage{amssymb}
\usepackage{cite}
\usepackage{graphicx}
\usepackage{array}
\usepackage{multirow}
\usepackage{amsmath}
\usepackage{stfloats}
\usepackage{graphicx}
\usepackage{subfigure}
\usepackage{tabularx}
\usepackage{booktabs}%
\usepackage{epsfig,epsf,balance,cite}
\usepackage[caption=false,font=normalsize,labelfont=sf,textfont=sf]{subfig}
\usepackage{setspace}
\usepackage{bm}
\usepackage{textcomp}
\usepackage{algorithmic}
\usepackage{algorithm}
\usepackage{subfigure}
\usepackage{caption}
\usepackage{graphicx}
\usepackage{setspace}
\usepackage{url}
\usepackage{verbatim}
\usepackage{graphicx}
\usepackage{cite}
\usepackage{amssymb}
\usepackage{color}
\usepackage{xpatch}
\definecolor{shadecolor}{RGB}{0,0,255}
\definecolor{blue}{RGB}{0,0,255}
\newtheorem{lemma}{Lemma}

\makeatletter

\ExplSyntaxOn
\cs_new:Npn \bibColoredItems #1#2
{
	\clist_map_inline:nn {#2} { \cs_new:cpn {bib@colored@##1} {#1} } 
}
\ExplSyntaxOff

\newcommand\bib@setcolor[1]{%
	\ifcsname bib@colored@#1\endcsname
	\expanded{\noexpand\color{\csname bib@colored@#1\endcsname}}%
	\else
	\normalcolor
	\fi
}

\xpatchcmd\@bibitem
{\item}
{\bib@setcolor{#1}\item}
{}{\fail}

\xpatchcmd\@lbibitem
{\item}
{\bib@setcolor{#2}\item}
{}{\fail}
\makeatother

\begin{document}
	
\title{Latency-Aware Resource Allocation for Integrated Communications, Computation, and Sensing in Cell-Free mMIMO Systems}
	
\author{Qihao Peng, Qu Luo, Zheng Chu, \\Zihuai Lin, ~\IEEEmembership{Senior Member,~IEEE}, Maged Elkashlan,~\IEEEmembership{Senior Member,~IEEE},\\
	Pei Xiao, ~\IEEEmembership{Senior Member,~IEEE}, George K. Karagiannidis, ~\IEEEmembership{Fellow,~IEEE} \\
	 and Christos Masouros,~\IEEEmembership{Fellow,~IEEE}.
		\thanks{Q. Peng, Q, Luo, and P. Xiao are affiliated with 5G and 6G Innovation Centre, Institute for Communication Systems (ICS) of University of Surrey, Guildford, GU2 7XH, UK. (e-mail: \{q.peng, q.u.luo, p.xiao\}@surrey.ac.uk). Z. Chu is with Department of Electrical and Electronic Engineering, University of Nottingham Ningbo China, Ningbo 315100, China. (e-mail: andrew.chuzheng7@gmail.com). Zihuai Lin is with the School of Electrical and Information Engineering, The University of Sydney, Sydney, NSW 2006, Australia. (e-mail: zihuai.lin@sydney.edu.au). M. Elkashlan is with the School of Electrical Engineering and Computer Science of Queen Mary University of London. (e-mail: m.elkashlan@qmul.ac.uk). George K. Karagiannidis is with Department of Electrical and Computer Engineering, Aristotle University of Thessaloniki, Greece. (e-mail: geokarag@auth.gr). Christos Masouros is with the Department of Electrical and Electronic Engineering, University College London, UK. (email: chris.masouros@ieee.org). (Corresponding author: Qu Luo)}} 
%	}
	
	\maketitle
	
\begin{abstract}
	In this paper, we investigate a  cell-free massive multiple-input and multiple-output (MIMO)-enabled integration communication, computation, and sensing (ICCS) system, aiming to minimize the maximum computation latency to guarantee the stringent sensing requirements. We consider a two-tier offloading framework, where each multi-antenna terminal can optionally offload its local tasks to either multiple mobile-edge servers for distributed computation or the cloud server for centralized computation while satisfying the sensing requirements and power constraint. The above offloading problem is formulated as a mixed-integer programming and non-convex problem, which can be decomposed into three sub-problems, namely, distributed offloading decision, beamforming design, and execution scheduling mechanism. First, the continuous relaxation and penalty-based techniques are applied to tackle the distributed offloading strategy. Then, the weighted minimum mean square error (WMMSE) and successive convex approximation (SCA)-based lower bound are utilized to design the integrated communication and sensing (ISAC) beamforming. Finally, the other resources can be judiciously scheduled to minimize the maximum latency. A rigorous convergence analysis and numerical results substantiate the effectiveness of our method. Furthermore, simulation results demonstrate that multi-point cooperation in cell-free massive MIMO-enabled ICCS significantly reduces overall computation latency, in comparison to the benchmark schemes.
\end{abstract}	
	

\section{Introduction}
Recently, integrated communication and sensing (ISAC) has emerged as one of the potential paradigms for the sixth generation (6G) wireless communication systems \cite{liu2022survey,liu2022integrated,zhang2021enabling}. ISAC is able to simultaneously achieve high-accuracy sensing and high-quality wireless connectivity, leveraging  the similarities in hardware and signal processing. Compared to the legacy wireless networks, the ISAC-enabled systems can significantly improve spectrum efficiency, reduce infrastructure costs, and offer flexible and versatile solutions for a wide range of applications, such as autonomous driving \cite{fu2021survey,cheng2022integrated} and smart manufacturing \cite{khan2020secured}.

The integration of ISAC with massive multiple-input multiple-output (MIMO) techniques offers significantly spatial beamforming gains, enhancing both communication and sensing capabilities \cite{liu2020joint,liao2024power}, and thus MIMO-ISAC has extracted extensive research attentions from academia and industry \cite{liu2021cramer,xiong2023fundamental,ren2023fundamental,he2023full,meng2024network,babu2024precoding, mao2024communication,elfiatoure2023cell}. For example, by optimizing the ISAC beamforming, the Cram{\'e}r-Rao bound (CRB) for sensing is derived in \cite{liu2021cramer} while considering the minimal communication requirement. Since echo and communication signals share the same frequency and time resources, they can significantly interfere with each other, ultimately degrading both communication and sensing performance. To theoretically analyze this issue, the authors of \cite{xiong2023fundamental} derive the CRB-rate region for a single target, revealing the trade-off bound between communication and sensing. Based on their results in \cite{xiong2023fundamental}, the fundamental trade-off between the achievable rate and the multi-target estimation CRB is analyzed in \cite{ren2023fundamental}. 
Considering the full-duplex operation for both radar and communication, the downlink precoding scheme and uplink transmission power are jointly devised to maximize the sum rate \cite{he2023full}. Then, to mitigate the severe inter-cell interference, the authors of \cite{meng2024network} proposed an interference management, and the beamforming is devised by exploiting the multi-cell cooperation \cite{babu2024precoding}. Considering the cell-free massive MIMO systems, the cooperative CRB-rate bound is studied in the cell-free massive MIMO systems \cite{mao2024communication}. Furthermore, to tackle the self-interference posed by ISAC, the authors of \cite{elfiatoure2023cell} assume that each AP can work as sensing or communication mode only, and devise the optimal operation mode of each AP by solving zero–one programming problem. Although the ISAC technique exhibits the tremendous advantages, how to enable real-time control according to the acquired sensing information remains elusive.

To address the above challenges, extensive contributions have been devoted to explore the seamless integration of communication, computation, and sensing (ICCS) technology in the MIMO systems \cite{qi2020integrated,li2023integrated,ding2022joint,liu2024joint,zhao2024multi}. Particularly, by using the superposition property of multiple access channels to aggregate massive data streams, the authors of \cite{qi2020integrated} and \cite{li2023integrated} demonstrate that the over-the-air computation (AirComp) technique can significantly enhance the communication performance and computation error, respectively. By fully exploiting the potential of mobile edge computing (MEC), the communication and computation resources are judiciously optimized to minimize the energy consumption \cite{ding2022joint}. Then, considering the cloud computation (CC), the three-tier computation framework is investigated \cite{liu2024joint}. Furthermore, to minimize the CRB of radar sensing while guaranteeing users' communication requirements, the semi-definite relaxation (SDR)-based alternating optimization and SDR-based successive convex approximation (SCA) algorithms are proposed in \cite{zhao2024multi}. 
Although the existing research in \cite{qi2020integrated,li2023integrated,ding2022joint,liu2024joint,zhao2024multi} has revealed that the MIMO-enabled ICCS can significantly enhance system performance, it is still worth investigating how to implement ICCS in the cell-free massive MIMO systems and characterizing the benefits of cooperative communication and computation by using distributed APs or servers. 

To date, existing results have demonstrated significant benefits through the integration of computation and communication in cell-free massive MIMO systems \cite{ke2020massive,wang2023task,li2024joint,yu2020non}. For example, the advantage of distributed computation and cooperative communication is explored in \cite{ke2020massive}. Based on MEC and CC, the multi-tier offloading strategy is investigated to minimize computation latency with perfect fronthaul link \cite{wang2023task}. To analyze the impact of limited fronthaul capacity on system performance, the balance between offloading decision and computation is studied in \cite{li2024joint}. Then, to further reduce the burden of information exchange in the cell-free massive MIMO systems, the non-orthogonal transmissions is investigated in \cite{yu2020non}. Whilst the aforementioned contributions have demonstrated the merits of task offloading and computation within a cell-free massive MIMO framework, the seamless integration of sensing alongside communication and computation remains an open research area, thereby prompting our exploration of the cell-free massive MIMO-enabled ICCS.

To the best of the authors’ knowledge, there is a paucity of work that has exploited the distributed computation offloading strategy in cell-free massive MIMO-enabled ICCS systems. To fully unleash the potential of distributed and centralized computations in cell-free massive MIMO systems, the offloading strategy, beamforming, and execution resources are jointly devised to minimize the maximum computation latency, while satisfying the sensing requirements, power constraints and computation capacity. The main contributions of this paper are summarized as follows: 
\begin{enumerate}
	\item  By exploring the ICCS in the cell-free massive MIMO systems, we propose a two-tier offloading strategy for the efficient implementation to unlock the potential for distributed communication and computation. Relying on the MIMO beamforming technique, each multi-antenna terminal can optionally offload its task to multiple APs for mobile edge computation or cloud sever for cloud computation. In this paper, we aim to minimize the maximum latency by jointly optimizing the offloading strategy, beamforming, execution frequency, and other resources, which is formulated as a mix-integer programming and non-convex problem.
	\item To tackle this problem, we decompose it into three sub-problems, namely, distributed offloading decision, beamforming design, and execution scheduling mechanism. Firstly, to address the mix-integer programming problem, we adopt the continuous relaxation and penalty techniques to obtain the solution of offloading strategy. Then, with the given offloading decision, the weighted minimum mean square error (WMMSE) method and a lower bound based on SCA method are applied to solve the non-convex problem. Finally, the execution frequency and other resources can be obtained by solving the convex problem. Furthermore, the convergence of proposed algorithm is mathematically proved.
	\item By strategically allocating computational resources and designing communication beamforming, the maximum latency, including the computation and computation delays, can be significantly reduced, facilitating seamless control. Furthermore, while distributed offloading offers potential advantages over centralized approaches, inter-user communication interference remains a critical bottleneck, which is validated by our simulation results. More importantly, numerical result demonstrate that our proposed method is superior over other benchmarks, which verifies the effectiveness of our proposed method.
\end{enumerate}

The remainder of the paper  is organized as follows. The system model is provided and a general latency is given in Section II. The optimization algorithm is formulated and designed in Section III. In Section IV, extensive numerical results are presented. Finally, our conclusions are drawn in Section V. 

\section{System Model}

\begin{figure}
	\centering
	\includegraphics[width=3.2in]{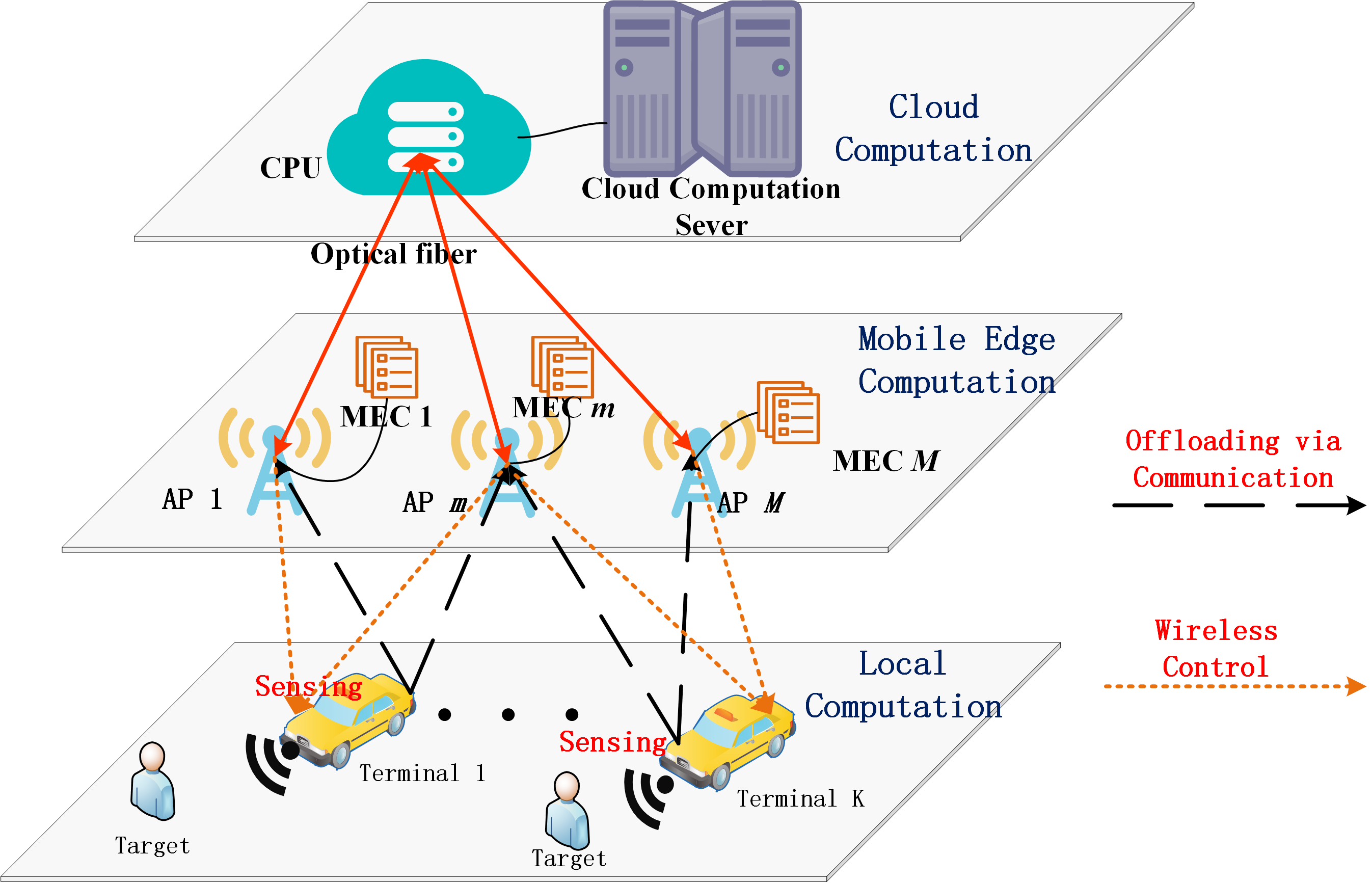}
	\caption{System model for ICCS in a cell-free mMIMO system.}
	\label{fig:system}
\end{figure}

As shown in Fig. \ref{fig:system}, \(M\) APs, each equipped with \(N\) antennas, are linked with an edge server to provide MEC service for $K$ mobile vehicles, each equipped with $N_t$ transmission antennas and \(N_r\) receiving antennas, respectively. Denoted the set of mobile vehicles and APs as \(\mathcal{K} = \{1,\cdots, K\} \) and \(\mathcal{M} = \{1,\cdots, M\} \), respectively. To enhance the system of scalability, we adopt the user-centric approach by defining \(\mathcal{M}_k \in \mathcal{M}\) as the set of APs that provide communication and computation service for the \(k\)-th terminal. In particular, the $k$-th mobile vehicle can sense the environment, including the location, velocity, and acceleration, to construct the data set with $D_k$ bits, $\forall k \in \mathcal{K}$. Mobile vehicles process computational resources and execute operations, such as braking and slowing down, in response to the concurrent sensing data. However, owing to the limited processing capacity, power consumption, and strict latency constraints, mobile vehicles are generally unable to process a large amount of data by relying on local computation. To tackle this issue, the \(k\)-th mobile vehicle can offload some specific tasks to either nearby APs for distributed computation or the centralized processing unit (CPU) for centralized computation. For MEC, the \(k\)-th vehicle can divide \(D_k\) bits into several parts, \(D_{m,k}\) bits with \(D_k = \sum \limits_{m \in \mathcal{M}_k} D_{m,k}\), and then transmit these tasks to several APs for distributed edge computation. Similarly, for CC, these tasks are first delivered to multiple APs, and then transmitted to the CPU for centralized cloud computation. 

%Besides, we assume that all mobile vehicles offload data based on TDMA to avoid severe interference. For one time slot, the $k$-th mobile car can offload $D_{k,m}$ bits, $\forall m \in \mathcal{M}_k$, to the $m$-th AP for computation.

\subsection{Edge-to-AP Communication Model}
The $k$-th mobile vehicle simultaneously transmits the signal \(s_{k,m}\) for offloading task to the \(m\)-th AP and radar signal \(s^{\text{sen}}_k\) for sensing, leading to the transmitted signal
\begin{equation}
	\label{transmitsignal}
	{\mathbf x}_k = \sum_{m \in \mathcal{M}_k} \mathbf{w}_{k,m} s_{k,m} + \mathbf{w}^{\text{sen}}_{k}\mathbf{s}^{\text{sen}}_k,
\end{equation}
where \( \mathbf{w}_{k,m} \in {\mathbb C}^{N_t \times 1} \) is the precoding vector designed for the $m$-th AP, and \(\mathbf{w}^{\text{sen}}_{k} \in \mathbb{C}^{N_t \times 1} \) represents the corresponding sensing beamforming. For ease of analysis but without loss of generality, we assume that the date symbol has unit power, i.e., \(\mathbb{E} \{s_{k,m}s^H_{k,m}\} = 1\). Furthermore, \(s_{k,m}\) and \(s^{\text{sen}}_k\) are assumed to be independent with each other.

The channel between the \(k\)-th vehicle and the \( m \)-th AP is defined as \(\mathbf{H}_{k,m} \in \mathbb{C}^{N \times N_t}\), and the \([n,m]\)-th element of \(\mathbf{H}_{k,m} \) is given by
\begin{equation}
	\label{elementofchannel}
	[\mathbf{H}_{k,m}]_{n,m} = \sqrt{\beta_{k,m}} h_{n,m},  \forall n \in [1, \cdots, N], \forall m \in [1,\cdots,N_t],
\end{equation} 
where \(\beta_{k,m}\) is the large-scale fading factor related to distance, and \(h_{n,m}\) is the small-scale fading factor that follows a complex Gaussian distribution with zero mean and unit variance. Then, the received signal \(\mathbf{y}_{k,m} \in {\mathbb{C}}^{N \times 1}\) at the \(m\)-th AP can be expressed as
\begin{equation}
	\label{receivedsignal}
	\begin{split}
		\mathbf{y}_{k,m} &= \sum^{K}_{k' = 1} \mathbf{H}_{k',m} \mathbf{x}_{k'} + \mathbf{n}_m, \\
		& = \mathbf{H}_{k,m}\mathbf{w}_{k,m}s_{k,m} + \mathbf{H}_{k,m} \Big( \sum_{m' \in \{\mathcal{M}_k \backslash m \}} \mathbf{w}_{k,m'} s_{k,m'} + \mathbf{w}^{\text{sen}}_{k}s^{\text{sen}}_k\Big)  \\
        &  \quad + \sum^{K}_{k' \neq k}{\sum_{m' \in \mathcal{M}_{k'}}  \Big(\mathbf{H}_{k',m}\mathbf{w}_{k',m'} s_{k',m'} + \mathbf{H}_{k',m}\mathbf{w}^{\text{sen}}_{k'}s^{\text{sen}}_{k'}\Big)} +  \mathbf{n}_m,
	\end{split}
\end{equation}
where \( \mathbf{n}_m \sim \mathcal{CN}(0, \mathbf{I}_N) \) represents the normalized additive white Gaussian noise (AWGN) at the \( m \)-th AP. Based on the received signal given in (\ref{receivedsignal}), the achievable data rate between the \(k\)-th vehicle and \(m\)-th AP can be expressed as \cite{1948Shanon}
\begin{equation}
	\label{datarate}
	R_{k,m} = B\log_2 \det\Big(\mathbf{I}_N+ \mathbf{H}_{k,m}\mathbf{w}_{k,m}\mathbf{w}^H_{k,m}\mathbf{H}^H_{k,m}{\mathbf{N}}^{-1}_{k,m}\Big),
\end{equation}
where $B$ is the bandwidth, $\mathbf{N}_{k,m}$ is the overall interference and background noise, which can be expressed as
\begin{equation}
	\label{noiseterm}
    \begin{split}
      {\mathbf{N}}_{k,m} &= \sum_{m' \in \{ {{\cal M}_k}\backslash m\} } \mathbf{H}_{k,m}\mathbf{w}_{k,m'}\mathbf{w}^H_{k,m'}\mathbf{H}^H_{k,m} +\mathbf{H}_{k,m}\mathbf{w}_{k}^{\text{sen}}(\mathbf{w}^\text{sen}_{k})^H\mathbf{H}^H_{k,m} \\
      &  \quad + \sum^{K}_{k' \neq k}{\sum_{m' \in \mathcal{M}_{k'}}  \Big(\mathbf{H}_{k',m}\mathbf{w}_{k',m'}\mathbf{w}^H_{k',m'}\mathbf{H}^H_{k',m} + \mathbf{H}_{k',m}\mathbf{w}^{\text{sen}}_{k'}(\mathbf{w}^{\text{sen}}_{k'})^H\mathbf{H}^H_{k',m}\Big)} + \mathbf{I}_N.  
    \end{split}
\end{equation}
As can be seen from (\ref{noiseterm}), inter-user interference escalates with an increasing number of offloading APs, leading to a decline in communication rates. Consequently, a trade-off exists between the number of offloading APs and the maximum latency.

%Finally, according to the Shannon capacity, the offloading latency from the \(k\)-th vehicle to the 
%\(m\)-th AP is given by
%\begin{equation}
%	\label{transmissionlatency}
%	T_{k,m}^{\text{tra}}  = \frac{D^c_{k,m}}{R_{k,m}},
%\end{equation}
%where \(D^c_{k,m}\) is the number of bits allocated by vehicle \(k\) to AP \(m\). 

\subsection{Edge-Node Sensing Model}
Next, we model the target echo signal at the vehicles' edge node , by assuming that the radar channel consists of line-of-sight (LoS) paths and both the transmit and receive uniform linear arrays (ULAs) at the vehicle are half-wavelength antenna spacing. Besides, the \(k\)-th vehicle can suppress the self-interference sufficiently via the corresponding techniques \cite{ding2022joint,liu2020jointMIMO}. The received echo signal is given by
\begin{equation}
	\label{echosignal}
	{\mathbf y}_k^s = \eta^{RT}_k \mathbf{a}_r(\theta_k)\mathbf{a}_t^H(\theta_k){\mathbf x}_k + \sum^K_{k'\neq k}{\mathbf{H}_{k,k'}{\mathbf x}_{k'}} + \mathbf{n}_s,
\end{equation}
where \(\eta_{RT} \) contains the impacts of the path
loss and complex reflection coefficients of the target and \(\theta_k\) represents the angle of arrival related to the \(k\)-th vehicle's sensing target \cite{liu2024joint}.  \(\mathbf{n}_s \sim \mathcal{CN}(0, \mathbf{I}_{N_r})\) denotes the normalized AWGN with zero mean and unit variance at the $k$-th vehicle, \(\mathbf{a}_r(\theta) = [1,e^{-j2\pi\frac{d}{\lambda}\sin \theta},\cdots,e^{-j2\pi\frac{d}{\lambda}(N_r -1)\sin \theta}]^T\) denotes the receive steering vector, and \(\mathbf{a}_t(\theta) = [1,e^{-j2\pi\frac{d}{\lambda}\sin \theta},\cdots,e^{-j2\pi\frac{d}{\lambda}(N_t -1)\sin \theta}]^T\) is transmitted array steering vector. \(\mathbf{H}_{k,k'} = \sqrt{\beta_{k,k'}} \mathbf{\Hat H}_{k,k'} \in \mathbb{C}^{N_r \times N_t} \) denotes the interference channel from vehicle \(k'\) to vehicle \(k\), \(\beta_{k,k'}\) is the large-scale fading factor related to distance, and \(\mathbf{\Hat H}_{k,k'}\) is the small-scale fading factors, each element of which follows the complex Gaussian distribution with zero mean and unit variance. Besides, we assume that  \(\eta^{RT}_k\) and \(\theta_k\) are known or previously coarse estimation at the vehicle for designing the best suitable transmit signal to detect this specific target of interest \cite{liu2024joint}. Then, based on (\ref{echosignal}), the $k$-th vehicle can capture the desired reflected signal of the point target, and thus the radar SINR can be written as
\begin{equation}
	\label{radSINR}
	\begin{split}
		\text{SINR}_{k}^{\text{sen}} &= \frac{\mathbb{E}\{\left\|\eta^{RT}_k\mathbf{a}_r(\theta_k)\mathbf{a}_t^H(\theta_k){\mathbf x}_k\right\|^2\}}{\mathbb{E}\{\|\mathbf{n}_s\|^2\} + \sum^K_{k'\neq k}{\mathbb{E}\{\|\mathbf{H}_{k,k'}{\mathbf x}_{k'}\|^2\}}} \\
		& = \frac{(\eta^{RT}_k)^2\text{tr}\{\mathbf{a}_r(\theta)\mathbf{a}_t^H(\theta_k)\mathbf{g}_k\mathbf{g}^H_k\mathbf{a}_t(\theta_k)\mathbf{a}_r^H(\theta_k)\} }{N_r +\sum^K_{k' \neq k} { \text{tr}\{\mathbf{H}_{k,k'}\mathbf{g}_k\mathbf{g}^H_k\mathbf{H}^H_{k,k'}\}}}, 
	\end{split}
\end{equation}
where \(\text{tr}\{\bf X\}\) denotes the trace of matrix \(\bf X\), $\mathbf{g}_k = \sum_{m \in \mathcal{M}_k} \mathbf{w}_{k,m} + \mathbf{w}^{\text{sen}}_{k} $ is sum of ISAC precoding vector from vehicle \(k\).
\subsection{Computation Model}
To characterize the computation strategy, i.e., local computation, mobile edge computation, and cloud computation, adopted by the \(k\)-th vehicle, \(\forall k \in \{1,\cdots, K\}\), we define the offloading variables \(\mathbf{b}_k = [b_{k,1},\cdots,b_{k,M}]^T\) with \(b_{k,m} \in [0,1]\), \(\forall m \in \{1,\cdots, M\}\) and \(\mathbf{c}_k = [c_{k,1},\cdots,c_{k,M}]^T\) with \(c_{k,m} \in [0,1]\), \(\forall m \in \{1,\cdots, M\}\). When \(b_{k,m} \neq 0 \) and \(c_{k,m} = 0\), \(\forall k, m\), the task is offloaded to the \(m\)-th edge server for computation. If \(b_{k,m} = 0 \) and \(c_{k,m} \neq 0\), \(\forall k, m\), the task is executed by the cloud computation server. For \(b_{k,m} = 0 \) and \(c_{k,m} = 0\), \(\forall k, m\), the task is performed by local vehicle \(k\). Based on this definition, we have
\begin{equation}
    \label{offload}
    \sum^{M}_{m=1} (b_{k,m} + c_{k,m}) \in \{0,1\}.
\end{equation}

\subsubsection{Local Computation}
For local computation, \(D_k\) bits need to be computed locally. Therefore, the local computation latency is given by
\begin{equation}
	\label{latencylocal}
	T^{\text {Loc}}_k = \frac{\alpha^\text {Loc}_k D_{k}}{f^\text {Loc}_k},
\end{equation}
where \(\alpha^\text {Loc}_k\) and \(f^\text {Loc}_k\) represent the local CPU cycle and CPU computing frequency of the vehicle \(k\), respectively. Then, the total power consumption of local computation and transmission is 
\begin{equation}
	\label{energylocal}
	P_k^{\text {Loc}} = \kappa_k^{\text {Loc}} (f^{\text {Loc}}_k)^3 + \mathbf{g}^H_k\mathbf{g}_k,
\end{equation}
where $\kappa_k^{\text {Loc}}$ is a constant related to the hardware architecture of vehicle \(k\).

\subsubsection{MEC}
For MEC, the \(k\)-th vehicle transmits \(D_{k}\) bits to the APs of set \(\mathcal{M}_k\). The transmission latency from the vehicle \(k\) to the \(m\)-th AP, \(\forall m \in \mathcal{M}_k\), is given by
\begin{equation}
	\label{transmissionlatency}
	T_{k,m}^{\text{MECTR}}  = \frac{b_{k,m}D_{k}}{R_{k,m}}.
\end{equation} 
Then, based on the received \(D_{k}\) bits, the \(m\)-th mobile edge server linked with the \(m\)-th AP can compute its decision. The edge computation latency is given by
\begin{equation}
	\label{latencymobile}
	T^{\text{MECc}}_{k,m} = \frac{\alpha^{\text{MEC}}_{m}b_{k,m} D_{k}}{f^{\text{MEC}}_{k,m}},
\end{equation}
where \(\alpha^{\text{MEC}}_{m}\) and \(f^{\text{MEC}}_{k,m}\) represent the CPU cycle and CPU computing frequency allocated to the \(k\)-th vehicle by AP \(m\), respectively. Based on the above discussions, the latency for processing tasks allocated by vehicle \(k\) to AP \(m\) can be expressed as
\begin{equation}
	\label{latencyDkm}
	T^{\text{MEC}}_{k} =\max \{T^{\text{MECTR}}_{k,m}+T^{\text{MECc}}_{k,m}\}, \forall m \in \mathcal{M}.
\end{equation}

Due to the limited computational capacity of the edge server, we have 
\begin{equation}
    \sum_{k \in \mathcal{K}} U(b_{k,m}) f^{\text{MEC}}_{k,m} \leq F^{\text{MEC},\max}_m, \forall m \in \mathcal{M},
\end{equation}
where  \(U(t)\) is 1 when \(t >0\) and \(U(t)\) is 0 for \(t \leq 0\).
Then, the power consumption of the \(m\)-th edge server is given by
\begin{equation}
    \label{MECPower}
    P^{\text{MEC}}_m = \sum_{k \in \mathcal{K}}{ U(b_{k,m}) \kappa_m^{\text {MEC}} (f^{\text {MEC}}_{k,m})^3}, \forall m \in \mathcal{M},
\end{equation}
where $\kappa_m^{\text {MEC}}$ is a constant related to the hardware architecture of the edge server \(m\).
%Then, the total latency for processing \(D_k\) bits is the maximum value between local computation latency and mobile edge computation latency, which can be expressed as
%\begin{equation}
%	\label{latencytotal}
%	T_{k} = \max \{T_k^{\text {loc}},T_{k,m}\}, \forall k \in [1,\cdots,K].
%\end{equation}
\subsubsection{CC}
The latency of cloud computation can be divided into three parts, including transmission latency from the vehicle to multiple APs, transmission latency from all APs to the CPU, and computation latency for cloud computation. Firstly, the \(k\)-th vehicle can simultaneously transmit data to multiple APs of set \(\mathcal{M}_k\), and the corresponding transmission latency is given by
\begin{equation}
    \label{ccU2APtransmissionlatency}
    T^{\text{V2ATra}}_{k,m} = \frac{c_{k,m}D_{k}}{R_{k,m}}, \forall m \in \mathcal{M}_k.
\end{equation}

Then, the received signal is delivered to the CPU linked to the cloud computation server. The transmission latency from the \(m\)-th AP to the CPU can be expressed as
\begin{equation}
    \label{ccA2Ctransmissionlatency}
    T^{\text{A2CTra}}_{k,m} = \frac{c_{k,m}D_{k}}{r^f_{k,m}}, \forall m \in \mathcal{M}_k,
\end{equation}
where \(r^f_{k,m}\) is the transmission bandwidth between the \(m\)-th AP and the CPU via optical fiber. 
Based on the received signal, the latency of cloud computation is given by
\begin{equation}
    \label{cclatency}
    T^{\text{CCom}}_{k,m} = \frac{\alpha^{\text{CC}}D_k}{f^{\text{CC}}_k},
\end{equation}
where \(\alpha^{\text{CC}}\) and \(f^{\text{CC}}_k\) are the processing density and the allocated local computation resource to the  \(k\)-th vehicle, respectively. 
Based on the above discussions, the total latency for cloud computation is given by
\begin{equation}
    \label{cctotallatency}
    T^{\text{CC}}_k = \max \{T^{\text{V2ATra}}_{k,m}+ T^{\text{A2CTra}}_{k,m} \} + T^{\text{CCom}}_{k,m},  \forall m \in \mathcal{M}_k.
\end{equation}

Due to the limited fronthaul capacity and computation capacity, we have
\begin{equation}
    \label{backhaul}
    \sum_{k \in \mathcal{K}} U(c_{k,m}) r^f_{k,m} \leq R^{f,\max}_m, \forall m \in \mathcal{M},
\end{equation}
and 
\begin{equation}
    \label{cccapacity}
    \sum_{k \in \mathcal{K}} (\sum_{m \in \mathcal{M}} c_{k,m}) f^{\text{CC}}_{k} \leq F^{\text{CC},\max},
\end{equation}
where \(R^{f,\max}_m\) is the fronthaul capacity from the \(m\)-th AP to the CPU and \(F^{\text{CC},\max}\) is the capacity for cloud computation. Furthermore, the power consumption of cloud computing is not taken into account, as cloud servers are typically supported by high-power infrastructure.

Based on the above discussions, the terminal \(k\) total latency can be expressed as \footnote{The results of computation can be transmitted by wireless communication, which is negligible compared to offloading and computation latency.}
\begin{equation}
    \label{totallatency}
    T^{\text{To}}_k = (1- \sum^{M}_{m=1} (b_{k,m} + c_{k,m}))T^{\text{loc}}_k + \sum^{M}_{m=1}b_{k,m} T^{\text{MEC}}_k + \sum^{M}_{m=1}c_{k,m} T^{\text{CC}}_k.
\end{equation}

\subsection{Problem Formulation}
Since the vehicle need to execute operations, including braking and slowing down, in response to the concurrent sensing information, how to achieve the seamless control will be a challenging problem. Therefore, we aim to minimize the maximum latency \( T^{\text{To}}_k \), \(\forall k \in \mathcal{K} \), by jointly optimizing the task allocation \(\mathbf{b}_k\), \(\mathbf{c}_k\), \(\forall k \in \mathcal{K} \), the communication beamforming vector \( \mathbf{w}_{k,m} \), the sensing beamformer \(\mathbf{w}_k^{\text {sen}}\), and execution frequency \(\{f^{\text{Loc}}_{k},f^{\text{MEC}}_{k,m},f^{\text{CC}}_{k}\}\), and transmission bandwidth \(r^f_{k,m}\), while satisfying the sensing requirement and power consumption. Mathematically, the problem can be formulated as

\begin{subequations}
	\label{Problem14}
	\begin{align}
	\min_{\substack{\{\mathbf{b}_k\}, \{\mathbf{c}_k\}, \{\mathbf{w}_{k,m}\}, \{\mathbf{w}_k^{\text{sen}}\} \\ \{f^{\text{Loc}}_{k}\},\{f^{\text{MEC}}_{k,m},\{f^{\text{CC}}_{k}\}, \{r^f_{k,m}\}}} \quad \max_{k \in \mathcal{K}} & \quad T^{\text{To}}_k \notag \\
	\text{s.t.} \quad & \sum^{M}_{m=1}b_{k,m},  \sum^{M}_{m=1}c_{k,m}, \sum^{M}_{m=1} (b_{k,m} + c_{k,m}) \in \{0,1\}, \forall k  \in \mathcal{K}, \label{Problema}\\
    & 0 \leq b_{k,m}, c_{k,m} \leq 1, \forall k  \in \mathcal{K}, \forall m \in \mathcal{M},\label{Problemb}\\
	& P_k^{\text{Loc}} \leq P^{\max}_{k}, \forall k  \in \mathcal{K}, \label{Problemc}\\
    & \sum_{k \in \mathcal{K}}{ U(b_{k,m}) \kappa_m^{\text {MEC}} (f^{\text {MEC}}_{k,m})^3} + \sum_{k \in \mathcal{K}}{ U(c_{k,m})}|\mathbf{H}_{k,m}\mathbf{w}_{k,m}|^2\notag \\
    & \quad  \leq P^{\text{MEC},\max}_m, \forall m \in \mathcal{M}, \label{Problemd}\\
    & \sum_{k \in \mathcal{K}} U(b_{k,m}) f^{\text{MEC}}_{k,m} \leq F^{\text{MEC},\max}_m, \forall m \in \mathcal{M}, \label{Probleme}\\
    & \sum_{k \in \mathcal{K}} U(c_{k,m}) r^f_{k,m} \leq R^{f,\max}_m, \forall m \in \mathcal{M},  \label{Problemf} \\
     & \sum_{k \in \mathcal{K}} (\sum_{m \in \mathcal{M}} c_{k,m}) f^{\text{CC}}_{k} \leq F^{\text{CC},\max}, \label{Problemg}  \\
	& \text{SINR}_{k}^{\text{sen}} \ge \text{SNR}_{k}^{\text{req}}, \forall k \in \mathcal{K},  \label{Problemh} 
    \end{align}
\end{subequations}
where constraints (\ref{Problema}) and (\ref{Problemb}) denote that the task offload strategy should satisfy the requirements, constraints (\ref{Problemc}) and (\ref{Problemd}) represent that the power consumption of each vehicle and each AP are both limited. Constraint (\ref{Probleme}) implies  that the computation capacity of each AP is limited, constraint (\ref{Problemf}) indicates the limited fronthaul link, constraint (\ref{Problemg}) denotes the limited computation capacity of cloud sever, and constraint (\ref{Problemh}) means that the required sensing accuracy should be satisfied. Problem (23) is not convex owing to the intractable form of mix-integrate constraint and non-convex constraints of (\ref{Probleme}), (\ref{Problemf}), and (\ref{Problemh}), which is challenging to solve.

\section{Algorithm Design} 
In this section, we propose an iteratively alternating optimization (AO) algorithm, which jointly adjusts the task offload strategy, optimizes the beamforming, and other resources to minimize the maximum latency of cell-free mMIMO systems while satisfying the sensing requirements and power budget.
\subsection{Problem Reforumlation}
First, by introducing a relax variable \(t\), Problem (\ref{Problem14}) can be equivalently transformed into
\begin{subequations}
	\label{Problem15}
	\begin{align}
	\min_{\substack{\{\mathbf{b}_k\}, \{\mathbf{c}_k\}, \{\mathbf{w}_{k,m}\}, \{\mathbf{w}_k^{\text{sen}}\} \\ \{f^{\text{Loc}}_{k}\},\{f^{\text{MEC}}_{k,m},\{f^{\text{CC}}_{k}\}, \{r^f_{k,m}\},t}} & \quad t \notag \\
	\text{s.t.} \quad & t \ge  (1-\sum_{m \in \mathcal{M}}{b_{k,m}} -\sum_{m \in \mathcal{M}}{c_{k,m}})T^{\text{Loc}}_k + \sum_{m \in \mathcal{M}}{b_{k,m}}T^{\text{MEC}}_k \notag \\
   & \quad +\sum_{m \in \mathcal{M}}{c_{k,m}}T^{\text{CC}}_k, \forall k \in \mathcal{K}, \label{Problem15a} \\
    & (\ref{Problema}), (\ref{Problemb}),(\ref{Problemc}),(\ref{Problemd}),(\ref{Probleme}),(\ref{Problemf}), (\ref{Problemg}), (\ref{Problemh}), \label{Problem15b}
	\end{align}
\end{subequations}
Owing to the intractable expression of (\ref{Problem15a}), Problem (\ref{Problem15}) is still challenging to solve. To tackle this issue, the AO algorithm is adopted to optimize the offloading strategy, beamforming, and other computation resources, respectively.

\subsection{Offloading Strategy Optimization}
With the given communication beamformer \(\{\mathbf{w}_{k,m}\}\), \(\forall k, m\), sensing beamformer  \(\{\mathbf{w}^{\text{sen}}_{k}\}\), \(\forall k\), and other resources \(\{f^{\text{Loc}}_{k}\},\{f^{\text{MEC}}_{k,m},\{f^{\text{CC}}_{k}\}, \{r^f_{k,m}\}\), Problem (\ref{Problem15}) can be solved by optimizing the offloading strategy \(\mathbf{b}_k\) and \(\mathbf{c}_k\), which can be equivalently written as
\begin{subequations}
	\label{AO1}
	\begin{align}
	\min_{\{\mathbf{b}_k\}, \{\mathbf{c}_k\},t} & \quad t \notag \\
	\text{s.t.} \quad & t \ge  (1-\sum_{m \in \mathcal{M}}{b_{k,m}} -\sum_{m \in \mathcal{M}}{c_{k,m}})T^{\text{Loc}}_k + \sum_{m \in \mathcal{M}}{b_{k,m}}T^{\text{MEC}}_k +\sum_{m \in \mathcal{M}}{c_{k,m}}T^{\text{CC}}_k, \forall k \in \mathcal{K}, \label{AO1a}\\
	%& \kappa_k^{\text {loc}} (f^{\text {loc}}_k)^3 \leq P^{\max}_{k} - \text{tr}\{\mathbf{V}_k\}, \forall k  \in \mathcal{K}, \label{AO1b}\\
      & \sum_{k \in \mathcal{K}}{ U(b_{k,m}) \kappa_m^{\text {MEC}} (f^{\text {MEC}}_{k,m})^3} + \sum_{k \in \mathcal{K}}{ U(c_{k,m})}|\mathbf{H}_{k,m}\mathbf{w}_{k,m}|^2 \leq P^{\text{MEC},\max}_m, \forall m \in \mathcal{M}, \label{AO1c}\\
    & (\ref{Problema}), (\ref{Problemb}), (\ref{Probleme}), (\ref{Problemf}), (\ref{Problemg}). \label{AO1e}
	\end{align}
\end{subequations}

As can be seen from (\ref{AO1}), it is still challenging to solve this problem as the constraint (\ref{Problem15a}) contains the maximum operation for latency. To address this issue, by introducing relax variables \(t^{\text{MEC}}_k\) and \(t^{\text{CC}}_k\), \(\forall k \in \mathcal{K}\), we obtain
\begin{subequations}
	\label{AO2}
	\begin{align}
	\min_{\{\mathbf{b}_k\}, \{\mathbf{c}_k\},\{t^{\text{MEC}}_k\},\{t^{\text{CC}}_k\},t} & \quad t \notag \\
	\text{s.t.} \quad & t \ge  (1-\sum_{m \in \mathcal{M}}{b_{k,m}} -\sum_{m \in \mathcal{M}}{c_{k,m}})T^{\text{Loc}}_k + \sum_{m \in \mathcal{M}}{b_{k,m}}t^{\text{MEC}}_k \notag \\
    & \quad +\sum_{m \in \mathcal{M}}{c_{k,m}}t^{\text{CC}}_k, \forall k \in \mathcal{K}, \label{AO2a}\\
    & t^{\text{MEC}}_k \ge \frac{b_{k,m}D_{k}}{R_{k,m}} + \frac{\alpha^{\text{MEC}}_{m}b_{k,m} D_{k}}{f^{\text{MEC}}_{k,m}}, \forall m \in \mathcal{M}, \label{AO2b} \\
    & t^{\text{CC}}_k \ge \frac{c_{k,m}D_{k}}{R_{k,m}} + \frac{c_{k,m}D_{k}}{r^f_{k,m}}+\frac{\alpha^{\text{CC}}D_k}{f^{\text{CC}}_k}, \forall m \in \mathcal{M}, \label{AO2c}\\
    &  (\ref{AO1c}), (\ref{Problema}), (\ref{Problemb}), (\ref{Probleme}), (\ref{Problemf}), (\ref{Problemg}). \label{AO2d}
	\end{align}
\end{subequations}

However, due to the binary constraints of \(\sum\limits_{m \in \mathcal{M}} b_{k,m}\) and \(\sum\limits_{m \in \mathcal{M}}c_{k,m}\), Problem (\ref{AO2}) is still intractable. To tackle this issue, Problem (\ref{AO2}) can be relaxed as
\begin{subequations}
	\label{AO3}
	\begin{align}
	\min_{\{\mathbf{b}_k\}, \{\mathbf{c}_k\},\{t^{\text{MEC}}_k\},\{t^{\text{CC}}_k\},t} & \quad t + \sum_{k \in \mathcal{K}}\rho^b_k(1-\mathbf{1}_{M \times 1}^T\mathbf{b}_k)\mathbf{1}_{M \times 1}^T\mathbf{b}_k+ \sum_{k \in \mathcal{K}}\rho^c_k(1-\mathbf{1}_{M \times 1}^T\mathbf{c}_k)\mathbf{1}_{M \times 1}^T\mathbf{c}_k\notag \\
	\text{s.t.} \quad & 0 \leq \sum_{m \in \mathcal{M}}(b_{k,m}+c_{k,m}) \leq 1, \forall k \in \mathcal{K}, \label{AO3b} \\
    & (\ref{AO2a}), (\ref{AO2b}), (\ref{AO2c}), (\ref{AO1c}), (\ref{Problemb}),(\ref{Probleme}), (\ref{Problemf}), (\ref{Problemg}), \label{AO3c}
	\end{align}
\end{subequations}
where \(\mathbf{1}_{M \times 1} \in \mathbb{N}^{M \times 1}\) is the vector of 1, \(\mathbf{b}_k = [b_{k,1},\cdots, b_{k,M}]^T\) and \(\mathbf{c}_k = [c_{k,1},\cdots, c_{k,M}]^T\) denote the vector of \(b_{k,m}\) and \(c_{k,m}\), respectively. \(\rho^b_k\) and \(\rho^b_k\) are the penalty factors. To solve this problem iteratively, the penalty factor in the  \(n\)-th iteration can be given by
\begin{equation}
    \label{penalty}
    \begin{split}
         \rho^{b,(n)}_k &= \upsilon (1-\mathbf{1}^T\mathbf{b}^{(n-1)}_k)\mathbf{1}^T\mathbf{b}^{(n-1)}_k, \\
         \rho^{c,(n)}_k &=  \upsilon(1-\mathbf{1}^T\mathbf{c}^{(n-1)}_k)\mathbf{1}^T\mathbf{c}^{(n-1)}_k,
    \end{split}
\end{equation}
where \(\mathbf{b}^{(n-1)}_k\) and \(\mathbf{c}^{(n-1)}_k\) are the solutions in the \((n-1)\)-th iteration.  \(\upsilon \) is a large constant that imposes the sum of \(b_{k,m}\) (or \(c_{k,m}\)) approaching 0 or 1. 

As can be seen from Problem (\ref{AO3}), the objective function and the constraints (\ref{AO2a}), (\ref{AO1c}), (\ref{Probleme}), and (\ref{Problemf}) are non-convex. Since the penalty term is a concave function with respect to \(\mathbf{b}_k\) or \(\mathbf{c}_k\), the penalty term can be upper bounded by
\begin{equation}
    \label{penaltyapp}
    G(\mathbf{b}_k) \leq (\mathbf{1}^T\mathbf{b}^{(n)}_k)+(1-2\times\mathbf{1}^T\mathbf{b}^{(n)}_k)\mathbf{1}^T\mathbf{b}_k.
\end{equation}
Then, the Taylor approximation is adopted to approximate the original function of (\ref{AO2a}), which is given by
\begin{equation}
    \label{xytaylor}
    t^{\text{MEC}}_k \mathbf{1}_{M \times 1}^T \mathbf{b}_k \approx \mathbf{1}_{M \times 1}^T \mathbf{b}^{(n)}_kt^{\text{MEC}}_k   + t^{\text{MEC},(n)}_k \mathbf{1}_{M \times 1}^T\mathbf{b}_k - t^{\text{MEC},(n)}_k \mathbf{1}_{M \times 1}^T \mathbf{b}^{(n)}_k.
\end{equation}
The term \(  \mathbf{1}_{M \times 1}^T \mathbf{c}_k t^{\text{CC}}_k\) can be approximated in a similar way. 
Finally, owing to the expression of \(U(\cdot)\), constraints (\ref{AO1c}), (\ref{Probleme}), and (\ref{Problemf}) are not convex functions. To tackle this issue, the weights \(w^{U_b}_{k,m}\) and \(w^{U_c}_{k,m}\), \(\forall k \in \mathcal{K}, \forall m \in \mathcal{M}\), are introduced to approximate \(U(b_{k,m})\) and \(U(c_{k,m})\) iteratively. Particularly, at the \(n\)-th iteration, \(w^{U_b,(n)}_{k,m}\) and \(w^{U_c,(n)}_{k,m}\) can be respectively expressed  as
\begin{equation}
    \label{ufactor}
    w^{U_b,(n)}_{k,m} = \frac{1}{b^{(n-1)}_{k,m} + \epsilon},  w^{U_c,(n)}_{k,m} = \frac{1}{c^{(n-1)}_{k,m} + \epsilon},
\end{equation} 
where \(\epsilon\) is the constant term that provides stability. At its core, the large value of \(w^{U_b}_{k,m}\) or \(w^{U_c,(n)}_{k,m}\) encourages \(b_{k,m}\) or \(c_{k,m}\) to zero.

Based on the above discussion, Problem (\ref{AO1}) can be solved by an iterative process, which can be formulated as
\begin{subequations}
	\label{AO4}
	\begin{align}
	\min_{\{\mathbf{b}_k\}, \{\mathbf{c}_k\},\{t^{\text{MEC}}_k\},\{t^{\text{CC}}_k\},t} & \quad t + \sum_{k \in \mathcal{K}}[\rho^{b,(n)}_kG(\mathbf{b}_k)+\rho^{c,(n)}_kG(\mathbf{c}_k)]\notag \\
	\text{s.t.} \quad & t \ge  (1-x^b_k -x^c_k)T^{\text{Loc}}_k -x^{b,(n)}_kt^{\text{MEC},(n)}_k + t^{\text{MEC},(n)}_kx^b_k    + x^{b,(n)}_kt^{\text{MEC}}_k \notag \\
    & \quad -x^{c,(n)}_kt^{\text{CC},(n)}_k + t^{\text{CC},(n)}_kx^c_k + x^{c,(n)}_kt^{\text{CC}}_k , \forall k \in \mathcal{K}, \label{AO4a} \\
    & \sum_{k \in \mathcal{K}} w^{U_b,(n)}_{k,m} b_{k,m} \kappa_m^{\text {MEC}} (f^{\text {MEC}}_{k,m})^3 + \sum_{k \in \mathcal{K}} w^{U_c,(n)}_{k,m} c_{k,m} |\mathbf{H}_{k,m}\mathbf{w}_{k,m}|^2 \notag \\ 
    & \quad \leq P^{\text{MEC},\max}_m, \forall m \in \mathcal{M}, \label{AO4b}\\
    & \sum_{k \in \mathcal{K}} w^{U_b,(n)}_{k,m}b_{k,m} f^{\text{MEC}}_{k.m} \leq F^{\text{MEC}}_m, \forall m \in \mathcal{M}, \label{AO4c}\\
    & \sum_{k \in \mathcal{K}} w^{U_c,(n)}_{k,m}c_{k,m} R^f_{k,m} \leq R^f_m, \forall m \in \mathcal{M}, \label{AO4d}\\
    & (\ref{AO2b}), (\ref{AO2c}), (\ref{Problemb}), (\ref{Problemg}), \label{AO4f}
	\end{align}
\end{subequations}
where \(x^b_k = \mathbf{1}_{M \times 1}^T\mathbf{b}_k\) and \(x^{c}_k = \mathbf{1}_{M \times 1}^T\mathbf{c}_k\) are the sum of \(b_{k,m}\) and \(c_{k,m}\), respectively. As can be seen from (\ref{AO4}), it is a linear problem that can be solved effectively by using CVX. 

To execute the algorithm for task offloading, it is necessary to find a feasible initial solution for Problem (\ref{AO4}). With the given beamformer and execution frequency, it is easy to obtain the latency for local computation \(T^{\text{Loc}}_k\), \(\forall k\), while considering the extreme case, i.e., all tasks are executed by terminals locally. Then, with the given beamformer, computation frequency, and bandwidth resources, the task offloading strategy \(\mathbf{b}_k\), \(\forall k\), can be obtained by solving the following \(K\) sub-problems: 
\begin{subequations}
	\label{initial1}
	\begin{align}
	\min_{\mathbf{b}_k,t^{\text{MEC}}_k,\eta^{\text{MEC}}} & \quad \eta^{\text{MEC}} +\sum_{k \in \mathcal{K}}\rho^{b,(n)}_kG(\mathbf{b}_k)  \notag \\
	\text{s.t.} \quad & \eta^\text{MEC} \ge  (1 -\sum_{m \in \mathcal{M}}b_{k,m} )T^{\text{Loc}}_k  + \sum_{m \in \mathcal{M}}b^{(n)}_{k,m} (t^{\text{MEC}}_k -t^{\text{MEC},(n)}_k ) + \sum_{m \in \mathcal{M}}b_{k,m} t^{\text{MEC},(n)}_k, \label{initial1a} \\
    & t^{\text{MEC}}_k \ge \frac{b_{k,m}D_{k}}{R_{k,m}} + \frac{\alpha^{\text{MEC}}_{m}b_{k,m} D_{k}}{f^{\text{MEC}}_{k,m}}, \forall m \in \mathcal{M}, \label{initial1b} \\
    & 0 \leq  \sum_{m \in \mathcal{M}}b_{k,m} \leq 1. \label{initial1d}
	\end{align}
\end{subequations}
Similarly, the task offloading strategy \(\mathbf{c}_k\), \(\forall k\), can be obtained by solving the following \(K\) sub-problems: 
\begin{subequations}
	\label{initial2}
	\begin{align}
		\min_{\mathbf{c}_k,t^{\text{CC}}_k,\eta^{\text{CC}}} & \quad \eta^{\text{CC}}+\sum_{k \in \mathcal{K}}\rho^{c,(n)}_kG(\mathbf{c}_k) \notag \\
		\text{s.t.} \quad & \eta^\text{CC} \ge  (1 -\sum_{m \in \mathcal{M}}c_{k,m} )T^{\text{Loc}}_k  + \sum_{m \in \mathcal{M}}c^{(n)}_{k,m} (t^{\text{CC}}_k -t^{\text{CC},(n)}_k ) + \sum_{m \in \mathcal{M}}c_{k,m} t^{\text{CC},(n)}_k, \label{initial1a} \\
		& t^{\text{CC}}_k \ge \frac{c_{k,m}D_{k}}{R_{k,m}} + \frac{c_{k,m}D_{k}}{r^f_{k,m}}+\frac{\alpha^{\text{CC}}D_k}{f^{\text{CC}}_k}, \forall m \in \mathcal{M}, \label{initial2b} \\
		& 0 \leq  \sum_{m \in \mathcal{M}}c_{k,m} \leq 1. \label{initial2c}
	\end{align}
\end{subequations}
After solving the above \(2K\) sub-problems, each terminal's latency can be minimized and the corresponding offloading strategy can be obtained. Based on the initial offloading strategy, the detailed algorithm is given in Algorithm \ref{algorithm1}.

\begin{algorithm}[t]
	\caption{Alternating Iterative Algorithm For Solving Problem (\ref{AO4})}
	\begin{algorithmic}[1]
		\label{algorithm1}
		\STATE The iteration number \(n\) and error tolerance $\zeta$ are initialized as 1 and 0.01, respectively;
		\STATE With the given beamformer  \(\mathbf{w}_{k,m}\), \(\mathbf{w}_k^{\text{sen}}\),  allocated execution frequency \(f^{\text{loc}}_k\),  \(f^{\text{MEC}}_{k,m}\),  \(f^{\text{CC}}_k\), and allocated bandwidth \(r^{f}_{k,m}\), \(\forall k,m\), solve Problem (\ref{initial1}) and Problem (\ref{initial2}) to respectively initialize the offloading strategy \(\mathbf{b}^{(n)}_k\) and  \(\mathbf{c}^{(n)}_k\), \(\forall k\), and calculate the total latency ${\rm{Obj}}^{\left(n\right)} = \max\limits_{k \in \mathcal{K}} T^{\text{To},(n)}_k$. Set ${\rm{Obj}}^{\left(0\right)} = 0$;
	%	\WHILE {${{\big( {{{\rm{Obj}}^{\left( n \right)}} - {{\rm{Obj}}^{\left( {n - 1} \right)}}} \big)} \mathord{\left/
		%			{\vphantom {{\left( {{{\rm{Obj}}^{\left( n \right)}} - {{\rm{Obj}}^{\left( {n - 1} \right)}}} \right)} {{{\rm{Obj}}^{\left( {n - 1} \right)}}}}} \right.
			%		\kern-\nulldelimiterspace} {{{\rm{Obj}}^{\left( {n - 1} \right)}}}} \ge \zeta$}
                    \WHILE {$\frac{{\rm Obj}^{(n)}-{\rm Obj}^{(n-1)}}{{\rm Obj}^{(n)}} >\zeta$}
                    \STATE Calculate \(\rho^{b,(n)}_k\), \(\rho^{c,(n)}_k\), \( w^{U_b,(n)}_{k,m}\), \( w^{U_c,(n)}_{k,m}\), latency for mobile computation \(t^{\text{MEC},(n)}_k\), and latency for cloud computation \(t^{\text{CC},(n)}_k\), based on the given offloading strategy \(\mathbf{b}^{(n)}_k\) and  \(\mathbf{c}^{(n)}_k\), \(\forall k\);
                    \STATE Update the iteration number by \(n = n + 1\);
                    \STATE Use the CVX tool to solve Problem (\ref{AO4}) and obtain the offloading strategy  \(\mathbf{b}^{(n)}_k\) and  \(\mathbf{c}^{(n)}_k\), \(\forall k\);
                    \STATE Calculate the total latency ${\rm{Obj}}^{\left(n\right)} = \max\limits_{k \in \mathcal{K}} T^{\text{To},(n)}_k$ with the given offloading strategy, beamformer, and execution frequency;
                    \ENDWHILE
		%\ENDWHILE
	\end{algorithmic}
\end{algorithm}

\subsection{Beamforming Optimization}

With the given task offloading strategy \(\mathbf{b}_k\), \(\mathbf{c}_k\), \(\forall k \in \mathcal{K}\), and and other resources \(\{f^{\text{Loc}}_{k}\}\),\(\{f^{\text{MEC}}_{k,m}\),\(\{f^{\text{CC}}_{k}\}\), \(\{r^f_{k,m}\}\), Problem (\ref{Problem14}) can be simplified to 
\begin{subequations}
	\label{AO5}
	\begin{align}
	\min_{ \{\mathbf{w}_{k,m}\}, \{\mathbf{w}_k^{\text{sen}}\},t} & \quad t \notag \\
	\text{s.t.} \quad & t \!\ge  \!(1\!-\!\sum_{m \in \mathcal{M}}{b_{k,m}} \!-\!\sum_{m \in \mathcal{M}}\!{c_{k,m}})T^{\text{Loc}}_k + \sum_{m \in \mathcal{M}}{b_{k,m}}T^{\text{MEC}}_k \! +\!\sum_{m \in \mathcal{M}}{c_{k,m}}T^{\text{CC}}_k, \forall k \in \mathcal{K}, \label{AO5a} \\
    &  \mathbf{g}^H_k\mathbf{g}_k \leq P^{\max}_{k} -\kappa_k^{\text {loc}} (f^{\text {Loc}}_k)^3, \forall k \in \mathcal{K}, \label{AO5b} \\
      &  \sum_{k \in \mathcal{K}}{ U(c_{k,m})}|\mathbf{H}_{k,m}\mathbf{w}_{k,m}|^2  \leq P^{\text{MEC},\max}_m \! -\! \sum_{k \in \mathcal{K}}{ U(b_{k,m}) \kappa_m^{\text {MEC}} (f^{\text {MEC}}_{k,m})^3}, \forall m \in \mathcal{M}, \label{AO5c}\\
    & \text{SINR}_{k}^{\text{sen}} \ge \text{SNR}_{k}^{\text{req}}, \forall k \in \mathcal{K}, \label{AO5d}
	\end{align}
\end{subequations}

Similarly, introducing the auxiliary variables \(t^{\text{MEC}}_k\) and \(t^{\text{CC}}_k\), \(\forall k \in \mathcal{K}\), Problem (\ref{AO5}) can be equivalently transformed into
\begin{subequations}
	\label{AO6}
	\begin{align}
	\min_{ \{\mathbf{w}_{k,m}\}, \{\mathbf{w}_k^{\text{sen}}\}, \{t^{\text{MEC}}_k\} , \{t^{\text{CC}}_k\},t} & \quad t \notag \\
	\text{s.t.} \quad & t \ge  (1-\sum_{m \in \mathcal{M}}{b_{k,m}} -\sum_{m \in \mathcal{M}}{c_{k,m}})T^{\text{Loc}}_k + \sum_{m \in \mathcal{M}}{b_{k,m}}t^{\text{MEC}}_k \notag \\
   & \quad +\sum_{m \in \mathcal{M}}{c_{k,m}}t^{\text{CC}}_k, \forall k \in \mathcal{K}, \label{AO6a} \\
   & t^{\text{MEC}}_k \ge \frac{b_{k,m}D_{k}}{R_{k,m}} + \frac{\alpha^{\text{MEC}}_{m}b_{k,m} D_{k}}{f^{\text{MEC}}_{k,m}}, \forall m \in \mathcal{M}, \label{AO6b} \\
    & t^{\text{CC}}_k \ge \frac{c_{k,m}D_{k}}{R_{k,m}} + \frac{c_{k,m}D_{k}}{R^f_{k,m}}+\frac{\alpha^{\text{CC}}D_k}{f^{\text{CC}}_k}, \forall m \in \mathcal{M}, \label{AO6c}\\
    & (\ref{AO5b}), (\ref{AO5c}), (\ref{AO5d}). \label{AO6d}
	\end{align}
\end{subequations}

However, Problem (\ref{AO6}) is still a non-convex problem due to the non-convex constraints (\ref{AO6b}), (\ref{AO6c}), and (\ref{AO5d}). To tackle this issue, we first rewrite the constraint (\ref{AO6b}) as
\begin{equation}
    \label{AO6btransform}
    R_{k,m} \ge \frac{b_{k,m}D_{k}}{(t^{\text{MEC}}_k-\frac{\alpha^{\text{MEC}}_{m}b_{k,m} D_{k}}{f^{\text{MEC}}_{k,m}})}, \forall m \in \mathcal{M}.
\end{equation}
To address this non-convex constraint, we adopt the iterative WMMSE-based method. By assuming that the \(m\)-th AP applies the linear receiver \(\mathbf{v}_{k,m} \in \mathbb{C}^{N \times 1}\) to detect the signal from the \(k\)-th terminal, the mean square error (MSE) can be expressed as
\begin{equation}
    \label{MSE}
    \begin{split}
       \text{MSE}_{k,m} &= \mathbb{E}\{(\mathbf{v}^H_{k,m}\mathbf{y}_{k,m} - s_{k,m})(\mathbf{v}^H_{k,m}\mathbf{y}_{k,m} - s_{k,m})^H\} \\
       & = 1+\mathbf{v}^H_{k,m}(\sum_{k'=1}^{K}\sum_{m \in \mathcal{M}_{k'}}\mathbf{H}_{k',m}\mathbf{w}_{k',m}\mathbf{w}^H_{k',m}\mathbf{H}^H_{k',m})\mathbf{v}_{k,m}- \mathbf{v}^H_{k,m}\mathbf{H}_{k,m}\mathbf{w}_{k,m} \\
       & \quad - \mathbf{w}^H_{k,m}\mathbf{H}^H_{k,m}\mathbf{v}_{k,m}+\mathbf{v}^H_{k,m}\mathbf{v}_{k,m} \\
       & = 1+\mathbf{v}^H_{k,m}\big(\mathbf{N}_{k,m} + \mathbf{H}_{k,m}\mathbf{w}_{k,m}\mathbf{w}^H_{k,m}\mathbf{H}_{k,m}\big)\mathbf{v}_{k,m}-\mathbf{v}^H_{k,m}\mathbf{H}_{k,m}\mathbf{w}_{k,m}-\mathbf{w}^H_{k,m}\mathbf{H}^H_{k,m}\mathbf{v}_{k,m}.
    \end{split}
\end{equation}
With the given beamformer \(\mathbf{w}_{k,m}\), the optimal linear detection vector \(\mathbf{v}_{k,m}\) can be obtained as
\begin{equation}
    \label{MMSEreceiver}
    \mathbf{v}^{\text{opt}}_{k,m} = \Big(\mathbf{N}_{k,m} + \mathbf{H}_{k,m}\mathbf{w}_{k,m}\mathbf{w}^H_{k,m}\mathbf{H}_{k,m}\Big)^{-1}\mathbf{H}_{k,m}\mathbf{w}_{k,m},
\end{equation}
and the corresponding MSE is given by
\begin{equation}
    \label{optMSE}
    \begin{split}
    	\text{MSE}^{\text{opt}}_{k,m} &= 1- \mathbf{w}^H_{k,m}\mathbf{H}^{H}_{k,m}\Big(\mathbf{N}_{k,m} + \mathbf{H}_{k,m}\mathbf{w}_{k,m}\mathbf{w}^H_{k,m}\mathbf{H}_{k,m}\Big)^{-1}\mathbf{H}_{k,m}\mathbf{w}_{k,m}.
    \end{split}
\end{equation}

By using the optimal detection vector \(\mathbf{v}_{k,m}\) and introducing auxiliary weight \(V_{k,m}\),\(\forall k,m\), we obtain the following lemma.
\begin{lemma}
\label{upperlemma}
The rate expression \(R_{k,m}\) can be equivalently written as
\begin{equation}
    \label{inequalitylem}
    R_{k,m} =  B \Big( \log_2(V_{k,m})-V_{k,m}\text{MSE}_{k,m}+1\Big),
\end{equation}
where the optimal \(V_{k,m}\) is given by
\begin{equation}
    \label{V_km}
    V^{\text{opt}}_{k,m} = \Big( 1 - \mathbf{w}^H_{k,m}\mathbf{H}^{H}_{k,m}\Big(\mathbf{N}_{k,m} + \mathbf{H}_{k,m}\mathbf{w}_{k,m}\mathbf{w}^H_{k,m}\mathbf{H}_{k,m}\Big)^{-1}\mathbf{H}_{k,m}\mathbf{w}_{k,m}\Big)^{-1}
\end{equation}
\textit{Proof}: Refer to Appendix A.  $\hfill\blacksquare$
\end{lemma}

With the given \(V_{k,m}\) and \(\mathbf{v}_{k,m}\), \(\forall k,m\), we rewrite the data rate between the \(k\)-th terminal and \(m\)-th AP as 
\begin{equation}
    \label{rwrate}
    \begin{split}
       R_{k,m}& = B\log_2(V_{k,m})-BV_{k,m}\mathbf{v}^H_{k,m}\big(\mathbf{N}_{k,m} + \mathbf{H}_{k,m}\mathbf{w}_{k,m}\mathbf{w}^H_{k,m}\mathbf{H}_{k,m}\big)\mathbf{v}_{k,m}\\
       & \quad+BV_{k,m}\mathbf{v}^H_{k,m}\mathbf{H}_{k,m}\mathbf{w}_{k,m}+BV_{k,m}\mathbf{w}^H_{k,m}\mathbf{H}^H_{k,m}\mathbf{v}_{k,m} + B - BV_{k,m}.
    \end{split}
\end{equation}
Obviously, it is a concave function with respect to \(\mathbf{w}_{k,m}\), \(\forall k, m\). By substituting the expression of (\ref{rwrate}) into (\ref{AO6b}), constraint (\ref{AO6b}) is a convex function.  
Similarly, constraint (\ref{AO6c}) can be equivalently transformed into 
\begin{equation}
    \label{AO6ctrans}
    \begin{split}
    	& B\log_2(V_{k,m})-BV_{k,m}-BV_{k,m}\mathbf{v}^H_{k,m}\big(\mathbf{N}_{k,m} + \mathbf{H}_{k,m}\mathbf{w}_{k,m}\mathbf{w}^H_{k,m}\mathbf{H}_{k,m}\big)\mathbf{v}_{k,m}\\
    	+& BV_{k,m}\mathbf{v}^H_{k,m}\mathbf{H}_{k,m}\mathbf{w}_{k,m}+BV_{k,m}\mathbf{w}^H_{k,m}\mathbf{H}^H_{k,m}\mathbf{v}_{k,m} + B \ge \frac{c_{k,m}D_{k}}{t^{\text{CC}}_k -\frac{c_{k,m}D_{k}}{R^f_{k,m}}-\frac{\alpha^{\text{CC}}D_k}{f^{\text{CC}}_k}  },  \forall m \in \mathcal{M}. 
    \end{split}
\end{equation}
Then, to address the non-convex constraint (\ref{AO5d}), we adopt the following lemma.
\begin{lemma}
	\label{lemma2}
	For a deterministic semi-positive matrix \(\mathbf{A} \in \mathbb{C}^{N_t\times N_t } \succeq \mathbf{ 0} \) and any vector $\mathbf{g}^{(n)}\in\mathbb{C}^{N_t \times 1}$, \(\mathbf{g}^H \mathbf{A}\mathbf{g} \) is lower bounded by
	\begin{equation}
		\mathbf{g}^H \mathbf{A}\mathbf{g} \ge 2({\mathbf{g}^{(n)}})^H \mathbf{A}\mathbf{g} -({\mathbf{g}^{(n)}})^H \mathbf{A}\mathbf{g}^{(n)},
	\end{equation}
where the equality holds only when \(\mathbf{g} = \mathbf{g}^{(n)}\).

\textit{Proof}: This proof can be readily obtained by using the Taylor expansion, which is omitted for brevity.  $\hfill\blacksquare$
\end{lemma}

By using Lemma (\ref{lemma2}), we have 
\begin{equation}
	\begin{split}
		&(\eta^{RT}_k)^2(\mathbf{g}^{(n)}_k)^H\mathbf{a}_t(\theta_k)\mathbf{a}_r^H(\theta_k)\mathbf{a}_r(\theta)\mathbf{a}_t^H(\theta_k)(2\mathbf{g}_k - \mathbf{g}^{(n)}_k)\\
		  \ge& \text{SINR}^{\text{req}}_k(N_r + \sum^K_{k' \neq k} \text{tr}\{\mathbf{g}^H_{k'}\mathbf{H}_{k,k'}\mathbf{H}^H_{k,k'}\mathbf{g}_{k'}\}),
	\end{split}
\end{equation}
where \(\mathbf{g}^{(n)}_k\) is the vector in the \(n\)-th iteration.

Problem (\ref{AO6}) is convex and can be readily solved by using CVX, which is detailed in Algorithm \ref{algorithm2}.

\begin{algorithm}[t]
	\caption{Alternating Iterative  Algorithm For Solving Problem (\ref{AO6})}
	\begin{algorithmic}[1]
		\label{algorithm2}
		\STATE The iteration number \(n\) and error tolerance $\zeta$ are initialized as 1 and 0.001, respectively;
		\STATE With the given offloading strategy \(\mathbf{b}_k\),  \(\mathbf{c}_k\), \(\forall k\),  allocated execution frequency \(f^{\text{loc}}_k\),  \(f^{\text{MEC}}_{k,m}\),  \(f^{\text{CC}}_k\), and allocated bandwidth \(r^{f}_{k,m}\), \(\forall k,m\), randomly initialize the beamformer \(\mathbf{w}^{(n)}_{k,m}\), \(\mathbf{w}_k^{\text{sen},(n)}\), \(\forall k, m\), and calculate the total latency ${\rm{Obj}}^{\left(n\right)} = \max\limits_{k \in \mathcal{K}} T^{\text{To},(n)}_k$. Set ${\rm{Obj}}^{\left(0\right)} = 0$;
	%	\WHILE {${{\big( {{{\rm{Obj}}^{\left( n \right)}} - {{\rm{Obj}}^{\left( {n - 1} \right)}}} \big)} \mathord{\left/
		%			{\vphantom {{\left( {{{\rm{Obj}}^{\left( n \right)}} - {{\rm{Obj}}^{\left( {n - 1} \right)}}} \right)} {{{\rm{Obj}}^{\left( {n - 1} \right)}}}}} \right.
			%		\kern-\nulldelimiterspace} {{{\rm{Obj}}^{\left( {n - 1} \right)}}}} \ge \zeta$}
                    \WHILE {$\frac{{\rm Obj}^{(n)}-{\rm Obj}^{(n-1)}}{{\rm Obj}^{(n)}} >\zeta$}
                    \STATE Calculate \( V^{(n)}_{k,m} \), \({\mathbf{v}}^{(n)}_{k,m}\), \(\mathbf{g}^{(n)}_k\), based on the given beamformer \(\mathbf{w}^{(n)}_{k,m}\), \(\mathbf{w}_k^{\text{sen},(n)}\), \(\forall k, m\);
                    \STATE Update the iteration number by \(n = n + 1\);
                    \STATE Use the CVX to solve Problem (\ref{AO6}) and obtain the beamformer \(\mathbf{w}^{(n)}_{k,m}\), \(\mathbf{w}_k^{\text{sen},(n)}\), \(\forall k, m\), ;
                    \STATE Calculate the total latency ${\rm{Obj}}^{\left(n\right)} = \max_{k \in \mathcal{K}} T^{\text{To},(n)}_k$ with the given offloading strategy, beamformer, and execution frequency;
                    \ENDWHILE
		%\ENDWHILE
	\end{algorithmic}
\end{algorithm}

\subsection{Execution Frequency Optimization}
With the given beamformer and offloading strategy, we aim to minimize the maximum latency by optimizing the execution frequency \(f^{\text{loc}}_k\),  \(f^{\text{MEC}}_{k,m}\),  \(f^{\text{CC}}_k\), and allocated bandwidth \(r^f_{k,m}\). Smililarly, by introducing the auxiliary variables \(t^{\text{MEC}}_k\) and \(t^{\text{CC}}\), \(\forall k \in \mathcal{K}\), problem can be formulated as
\begin{subequations}
	\label{AO7}
	\begin{align}
	\min_{\{f^{\text{loc}}_k\}, \{f^{\text{MEC}}_{k,m}\}, \{f^{\text{CC}}_k\}, \{r^f_{k,m}\}} \quad & \quad t \notag \\
	\text{s.t.} \quad&t \ge  (1-\sum_{m \in \mathcal{M}}{b_{k,m}} -\sum_{m \in \mathcal{M}}{c_{k,m}})T^{\text{Loc}}_k + \sum_{m \in \mathcal{M}}{b_{k,m}}t^{\text{MEC}}_k \notag \\
   & \quad +\sum_{m \in \mathcal{M}}{c_{k,m}}t^{\text{CC}}_k, \forall k \in \mathcal{K}, \label{AO7a}\\
      & t^{\text{MEC}}_k \ge \frac{b_{k,m}D_{k}}{R_{k,m}} + \frac{\alpha^{\text{MEC}}_{m}b_{k,m} D_{k}}{f^{\text{MEC}}_{k,m}}, \forall m \in \mathcal{M}, \label{AO7b} \\
    & t^{\text{CC}}_k \ge \frac{c_{k,m}D_{k}}{R_{k,m}} + \frac{c_{k,m}D_{k}}{f^f_{k,m}}+\frac{\alpha^{\text{CC}}D_k}{f^{\text{CC}}_k}, \forall m \in \mathcal{M}, \label{AO7c}\\
    & (\ref{Problemc}), (\ref{Problemd}), (\ref{Probleme}), (\ref{Problemf}), (\ref{Problemg}). \label{AO7d}
   % &\kappa_k^{\text {loc}} (f^{\text {loc}}_k)^3 \leq P^{\max}_{k} -\text{tr}\{\mathbf{V}_k\}, \forall k \in \mathcal{K}, \label{AO7b}\\
    %&  \sum_{k \in \mathcal{K}}{ U(b_{k,m}) \kappa_m^{\text {MEC}} (f^{\text {MEC}}_{k,m})^3}  \leq P^{\text{MEC},\max}_m \notag \\
    %& \quad -\sum_{k \in \mathcal{K}}{ U(c_{k,m})}|\mathbf{H}_{k,m}\mathbf{w}_{k,m}|^2, \forall m \in \mathcal{M},\label{AO7c}\\
    %& \sum_{k \in \mathcal{K}} U(b_{k,m}) f^{\text{MEC}}_{k,m} \leq F^{\text{MEC}}_m, \forall m \in \mathcal{M}, \label{AO7d}\\
    %& \sum_{k \in \mathcal{K}} U(c_{k,m}) R^f_{k,m} \leq R^f_m, \forall m \in \mathcal{M},  \label{AO7e} \\
     %& \sum_{k \in \mathcal{K}} (\sum_{m \in \mathcal{M}} c_{k,m}) f^{\text{CC}}_{k} \leq F^{\text{CC}}, \label{AO7f}
    \end{align}
\end{subequations}
This problem can be readily solved by using CVX.
%\begin{lemma}
 %   \label{lemma1}
  %  Given the \(x > 0\) and \(y >0\), the \(\frac{1}{\log_2(1+x/y)}\) is lower bounded by
   % \begin{equation}
    %    \frac{1}{\log_2(1+x/y)} \ge 
    %\end{equation}
    
    %\textit{Proof}: Please refer to Appendix A. \(\hfill \blacksquare\)
%\end{lemma}

%Owing to the transmission rate following the concavity of function \(\log_2 (1+z)\), we have the following inequality:
%\begin{equation}
 %   \label{lnxy}
  %  \log_2(1+\frac{x}{y}) \leq \log_2(1+\frac{x_0}{y_0}) + \frac{\ln{2}}{1+x_0/y_0} \times (\frac{0.5x^2/x_0 + 0.5x_0}{y}-\frac{x_0}{y_0}),
%\end{equation}
%where 

Based on the above discussions, the solutions for task allocation, ISAC beamforming, and execution frequency can be obtained by iteratively searching the solution for three optimization problems.

\subsection{Algorithm Analysis}
Next, we aim to analyze the convergence and complexity of our proposed algorithm in cell-free mMIMO systems.

\subsubsection{Convergence Analysis} Before proving the convergence of our proposed algorithms, one needs to prove that the solution in the \((n-1)\)-th iteration is also feasible for the \(n\)-th iteration. For the offloading strategy, owing to the simple linear programming and Taylor approximation, the offloading strategy and frequency execution can converge to the sub-optimal, while simultaneously satisfying \(t^{(n)} \leq t^{(n-1)}\). Therefore, we focus on the convergence of beamforming optimation. By denoting the \(n\)-th iteration's optimal solution of beamformer as \(\mathbf{w}^{(n)}_{k,m}\) and \(\mathbf{w}^{\text{sen},(n)}_k\), \(\forall k \in \mathcal{K}\), \(\forall m \in \mathcal{M}\), we have
\begin{equation}
    \label{convergence1}
    \begin{split} &(\eta^{RT}_k)^2\Big((\mathbf{g}^{(n-1)}_k)^H\mathbf{a}_t(\theta_k)\mathbf{a}_r^H(\theta_k)\mathbf{a}_r(\theta)\mathbf{a}_t^H(\theta_k)(2\mathbf{g}^{(n)}_k-\mathbf{g}^{(n-1)}_k)\Big)\\
    \ge& \text{SNR}^{\text{req}}_k(N_r + \sum^K_{k' \neq k} \text{tr}\{(\mathbf{g}^{(n)}_{k'})^H\mathbf{H}_{k,k'}\mathbf{H}^H_{k,k'}\mathbf{g}^{(n)}_{k'}\}),
    \end{split}
\end{equation}
where \(\mathbf{g}^{(n)}_k = \sum\limits_{m \in \mathcal{M}_k}\mathbf{w}^{(n)}_{k,m} +\mathbf{w}^{\text{sen},(n)}_k\) is the optimal solution in the \(n\)-th iteration. Based on  Lemma \ref{lemma2}, we have 
\begin{equation}
    \label{convergence2}
    \begin{split}
        &(\eta^{RT}_k)^2\Big((\mathbf{g}^{(n)}_k)^H\mathbf{a}_t(\theta_k)\mathbf{a}_r^H(\theta_k)\mathbf{a}_r(\theta)\mathbf{a}_t^H(\theta_k)\mathbf{g}^{(n)}_k\Big)\\
        \ge &(\eta^{RT}_k)^2\Big((\mathbf{g}^{(n-1)}_k)^H\mathbf{a}_t(\theta_k)\mathbf{a}_r^H(\theta_k)\mathbf{a}_r(\theta)\mathbf{a}_t^H(\theta_k)(2\mathbf{g}^{(n)}_k-\mathbf{g}^{(n-1)}_k)\\
        \ge & \text{SNR}^{\text{req}}_k(N_r + \sum^K_{k' \neq k} \text{tr}\{(\mathbf{g}^{(n)}_{k'})^H\mathbf{H}_{k,k'}\mathbf{H}^H_{k,k'}\mathbf{g}^{(n)}_{k'}\}).
    \end{split}
\end{equation}
Therefore, the optimal beamformer in the \((n-1)\)-th iteration is also feasible in the \(n\)-th iteration. Furthermore, based on the WMMSE-based method, \(\forall k \in \mathcal{K}\), \(\forall m \in \mathcal{M}\), we have 
\begin{equation}
	\begin{split}
		R^{(n)}_{k,m} &= B\log_2(V^{(n)}_{k,m}) - BV^{(n)}_{k,m}-BV^{(n)}_{k,m}(\mathbf{v}^{(n)}_{k,m})^H\big(\mathbf{N}_{k,m} + \mathbf{H}_{k,m}\mathbf{w}^{(n)}_{k,m}(\mathbf{w}^{(n)}_{k,m})^H\mathbf{H}_{k,m}\big)\mathbf{v}^{(n)}_{k,m}\\
		& \quad+2BV^{(n)}_{k,m}(\mathbf{v}^{(n)}_{k,m})^H\mathbf{H}_{k,m}\mathbf{w}^{(n)}_{k,m}(\mathbf{w}^{(n)}_{k,m})^H\mathbf{H}^H_{k,m}\mathbf{v}^{(n)}_{k,m} + B \\
		&\ge B\log_2(V^{(n-1)}_{k,m})-BV^{(n-1)}_{k,m}-BV^{(n-1)}_{k,m}(\mathbf{v}^{(n-1)}_{k,m})^H\big(\mathbf{N}_{k,m} + \mathbf{H}_{k,m}\mathbf{w}^{(n)}_{k,m}(\mathbf{w}^{(n)}_{k,m})^H\mathbf{H}_{k,m}\big)\mathbf{v}^{(n-1)}_{k,m}\\
		& \quad+2BV^{(n-1)}_{k,m}(\mathbf{v}^{(n-1)}_{k,m})^H\mathbf{H}_{k,m}\mathbf{w}^{(n)}_{k,m}(\mathbf{w}^{(n)}_{k,m})^H\mathbf{H}^H_{k,m}\mathbf{v}^{(n-1)}_{k,m} + B\ge R^{(n-1)}_{k,m}
	\end{split}
\end{equation}
Then, by substituting \(R^{(n)}_{k,m} \ge R^{(n-1)}_{k,m}\) into (\ref{AO6b}), we obtain  
\begin{equation}
    \label{MECn}
    \begin{split}
           &\frac{b_{k,m}D_{k}}{R^{(n)}_{k,m}} + \frac{\alpha^{\text{MEC}}_{m}b_{k,m} D_{k}}{f^{\text{MEC}}_{k,m}} \leq t^{\text{MEC},(n)}_k  \leq  \frac{b_{k,m}D_{k}}{R^{(n-1)}_{k,m}} + \frac{\alpha^{\text{MEC}}_{m}b_{k,m} D_{k}}{f^{\text{MEC}}_{k,m}} \leq t^{\text{MEC},(n-1)}_k.
    \end{split}
\end{equation}
It is readily obtained that \(t^{\text{MEC},(n)}_k \leq t^{\text{MEC},(n-1)}_k\), \(\forall k\). In a  similar way, we can prove that \(t^{\text{CC},(n)}\) is no larger than \(t^{\text{CC},(n-1)}\) based on the constraint (\ref{AO6c}). Finally, based on the above discussions, we have
\begin{equation}
    \label{converge3}
    \begin{split}
          t^{(n-1)} &\ge (1-\sum_{m \in \mathcal{M}}b_{k,m}-\sum_{m \in \mathcal{M}}c_{k,m})T^{\text{Loc},(n-1)}_k +  \sum_{m \in \mathcal{M}}b_{k,m}t^{\text{MEC},(n-1)}_k + \sum_{m \in \mathcal{M}}c_{k,m}t^{\text{CC},(n-1)}\\
          &\ge (1-\sum_{m \in \mathcal{M}}b_{k,m}-\sum_{m \in \mathcal{M}}c_{k,m})T^{\text{Loc},(n)}_k +  \sum_{m \in \mathcal{M}}b_{k,m}t^{\text{MEC},(n)}_k + \sum_{m \in \mathcal{M}}c_{k,m}t^{\text{CC},(n)}  \ge t^{(n)}.
    \end{split}
\end{equation}
Therefore, the convergence of our proposed algorithm is verified.
%Furthermore, we can prove that Algorithm can converge to the Karush-Kuhn-Tucker (KKT) point of Problem  for the abovementioned precoding schemes by using the similar proof as in Appendix B [].
\subsubsection{Complexity Analysis} The complexity for solving Problem (\ref{Problem14}) depends on the number of iterations and the complexity of each algorithm. Specifically, the main complexity of each iteration in Algorithm 1 lies in solving Problem (\ref{AO1})  which includes $(2 \sum_{k \in \mathcal{K}}|\mathcal{M}_k|K + 1)$ variables, where \(|\mathcal{M}_k|\) is the cardinality of set \(\mathcal{M}_k\). Therefore, the computational complexity of this algorithm is on the order of ${\mathcal{O}}\Big(N^{\text{iter}}_1 \times (2 \sum_{k \in \mathcal{K}}|\mathcal{M}_k|K + 1)^3 \Big)$, where $N^{\text{iter}}_1$ is the number of iterations. Similarly, the complexity for solving Problem (\ref{AO7}) is \(\mathcal{O}\Big( (5M+2K + 1)^3\Big) \). For Algorithm 2, the complexity mainly comes from the updating of \(\mathbf{w}_{k,m}\) and \(\mathbf{w}^{\text{sen}}_k\). Based on \cite{boyd2004convex}, the complexity for solving \(\mathbf{w}_{k,m}\) is \(\mathcal{O}\Big(\sum_{k \in \mathcal{K}}(|\mathcal{M}_k|+1) N^3_t\Big)\), and thus the complexity of Algorithm 2 is \({\mathcal{O}}\Big(\sum_{k \in \mathcal{K}}(|\mathcal{M}_k|+1)N^3_tN^{\text{iter}}_2\Big)\), where \(N^{\text{iter}}_2\) is the number of iterations. Therefore, the overall complexity of our proposed method is given by
\begin{equation}
	 \mathcal{O}\Big( N_{\text{iter}_1} \times (2 \sum_{k \in \mathcal{K}}|\mathcal{M}_k|K + 1)^3 +\sum_{k \in \mathcal{K}}(|\mathcal{M}_k|+1)N^3_tN^{\text{iter}}_2+ (5M+2K + 1)^3\Big).
\end{equation}

\section{Simulation Results} 
This section presents numerical results that validate the effectiveness of our proposed algorithm and demonstrate the performance improvement achieved by comparing our algorithm with other benchmarks.
\subsection{Simulation Setup}
$M$ APs are assumed to be randomly deployed in a $\rm 0.2$ km $\times$ $\rm 0.2$ km square, and then provide service for \(K\) randomly distributed terminals. Each terminal's sensing target is randomly distributed with the direction and distance generated by following a uniform distribution over \([0,\pi]\) and [40m,50m], and their reflection coefficients are assumed to be uniformly distributed between [0.8,1]. The large-scale fading coefficient model is adopted \cite{2017small}, which is given by
\begin{equation}
	\setlength\abovedisplayskip{5pt}
	\setlength\belowdisplayskip{5pt}
	\label{channel_model}
	{\rm{P}}{{\rm{L}}_{m,k}} = \left\{ {\begin{array}{*{20}{l}}
			{\begin{array}{*{20}{l}}
					{L_{\rm{loss}} \!+\! 35{{\log }_{10}}\left( {{d_{m,k}}} \right),{{d_{m,k}} \!>\! {d_1}}, }\\
					{L_{\rm{loss}} \!+\! 15{{\log }_{10}}\left( {{d_1}} \right) \!+\! 20{{\log }_{10}}\left( {{d_0}} \right),{{d_{m,k}} \!\le\! {d_0}},} \\
					{L_{\rm{loss}} \!+\! 15{{\log }_{10}}\left( {{d_1}} \right) \!+\! 20{{\log }_{10}}\left( {{d_{m,k}}} \right), \rm{other},}\\
			\end{array}}
	\end{array}} \right.
\end{equation}
where $d_{m,k} \left(\rm{km}\right)$ is the distance between the $m$-th AP and the $k$-th device, and $L_{\rm{loss}} \left(\rm{dB}\right)$ is a constant factor that depends on the carrier frequency $f\left(\rm{MHz}\right)$, the heights of the APs $h_{{\rm{AP}}} \left(\rm{m}\right)$ and devices $h_u \left(\rm{m}\right)$. Specifically, $L_{\rm {loss}}$ is given by
\begin{equation}
\setlength\abovedisplayskip{5pt}
\setlength\belowdisplayskip{5pt}
\label{path_loss_L}
\begin{split}
L_{\rm{loss}} &= 46.3 + 33.9{\log _{10}}\left( f \right) - 13.82{\log _{10}}\left( {{h_{{\rm{AP}}}}} \right) \\
 & - \left( {1.1{{\log }_{10}}\left( f \right) - 0.7} \right){h_u} + \left( {1.56{{\log }_{10}}\left( f \right) - 0.8} \right).
\end{split}
\end{equation}
For the small-scale fading factors, it is generally modeled as Rayleigh fading with zero mean and unit variance.
The noise power is related to bandwidth \(B\), Boltzmann constant  $k_B = 1.381 \times 10^{-23}$ (Joule per Kelvin), and temperature ${T_0} = 290$ (Kelvin), i.e.,
\begin{equation}
\setlength\abovedisplayskip{5pt}
\setlength\belowdisplayskip{5pt}
\label{noise_power}
P_n = B \times {k_B} \times {T_0} \times  10^{\frac{N_{\rm{dB}}}{10}} \left( {\rm{W}} \right).
\end{equation}
Unless otherwise stated, most of the simulation parameters are listed in Table \ref{tabel}. By using the user-centric approach, the APs of \(\mathcal{M}_k\) can be selected based on the large-scale fading factors in descending order \cite{peng2022resource}. Furthermore, we also provide the performance of the following three different schemes under the same parameter configuration.
\begin{itemize}
    \item \textbf{Local Scheme}: Each terminal can process the data locally without any information exchange, and thus the latency can obtained by jointly optimizing the sensing beamformer and local computation frequency.
    \item \textbf{MEC Scheme}: Since each terminal transmits the data to the nearby APs for mobile edge computation, the optimal latency can be achieved by jointly optimizing the beamformer and mobile computation resources.
    \item \textbf{CC Scheme}: Each terminal transmits the data to the CPU for cloud computation. Therefore, we jointly optimize the beamformer, bandwidth, and cloud computation frequency to minimize the maximum latency.
%    \item \textbf{Joint MEC and Local Scheme}: Each terminal can choose to offload the tasks to the MEC server or process them locally. As a result, the minimization of maximum latency can be solved by tailoring local and mobile resources. 
\end{itemize}

\begin{table}[t]
    \centering
     \caption{Simulation Parameters}
    \label{tabel}
    \begin{tabular}{cc}
    \toprule
    \textbf{Parameters} & Value\\
    \midrule
    Number of APs \(M\)  &  6\\
    Number of AP's Antennas \(N\) &8\\
    Number of Terminals \(K\)     & 6\\ 
    Number of Terminal's Transmission Antennas \(N_t\)  &8 \\ 
    Number of Terminal's Receiving Antennas \(N_r\)      & 8\\
    Transmission Bandwidth \(B\)    & 10 MHz\\
    Task Computation Intensity \(\alpha^{\text{Loc}}_k,\alpha^{\text{MEC}}_{k,m},\alpha^{\text{CC}}, \forall k, m\)    & 400 cycles/bit \\
    Numbers of Bits for Computation \(D_k, \forall k\)   & 0.2 MB \\
    Terminal's local Computation Frequency \(f^{\text{Loc}}_k,\forall k\)    & \(3\times 10^8\) cycles/s\\
    Mobile Computation Frequency  \(F^{\text{MEC}}_m,\forall m\)     & \(3\times 10^9\) cycles/s \\
    Cloud Computation Frequency   \(F^{\text{CC},\max}\)  &  \(1\times 10^{10}\) cycles/s\\
    fronthual bandwidth from AP to the CPU \(R^{f,\max}_m, \forall m\) & 0.5 GHz\\
    Constant of Hard Architecture \(\kappa^{\text{Loc}}_k,\kappa^{\text{MEC}}_m,\kappa^{\text{CC}}, \forall k, m\) & \(1 \times 10^{-28}\) \\
    Sensing SINR Requirement \(\text{SINR}^{\text{req}}_k\) & 1 dB \\
    Terminal's Transmission Power \(P^{\max}_k, \forall k\) & 23 dBm \\
    AP's Power \(P^{\text{MEC},\max}_m, \forall m\) & 30 dBm\\ 
    \bottomrule
    \end{tabular}
\end{table}

\subsection{Convergence of Proposed Algorithm}

\begin{figure}
	\centering
	\includegraphics[width=3.2in]{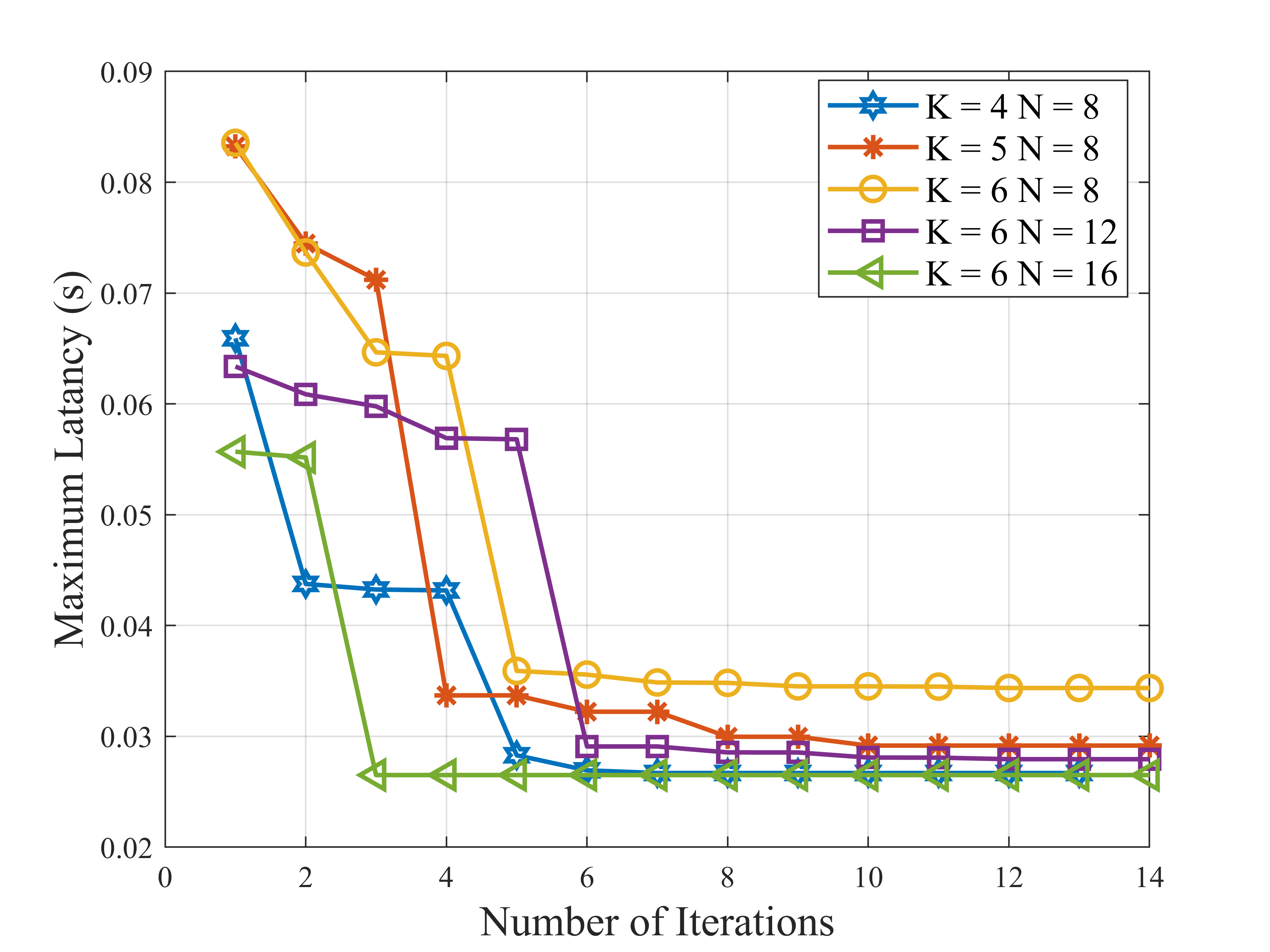}
	\caption{Convergence of our proposed algorithm with \(|\mathcal{M}_k| = 3, \forall k\).}
	\label{convergence}
\end{figure}

To verify the effectiveness of our proposed algorithm, we depict the maximum latency of each iteration in Fig. \ref{convergence}.  The proposed method converges to a local minimum solution rapidly within 10 iterations with various parameter settings. It is worth noting that an increase in the  number of APs' antennas can decrease the maximum latency owing to the enhanced receiving gain. Furthermore, the increasing number of terminals will significantly deteriorate the system performance due to increased interference.

\subsection{Effect of Number of Serving APs}

\begin{figure}
	\centering
	\includegraphics[width=3.2in]{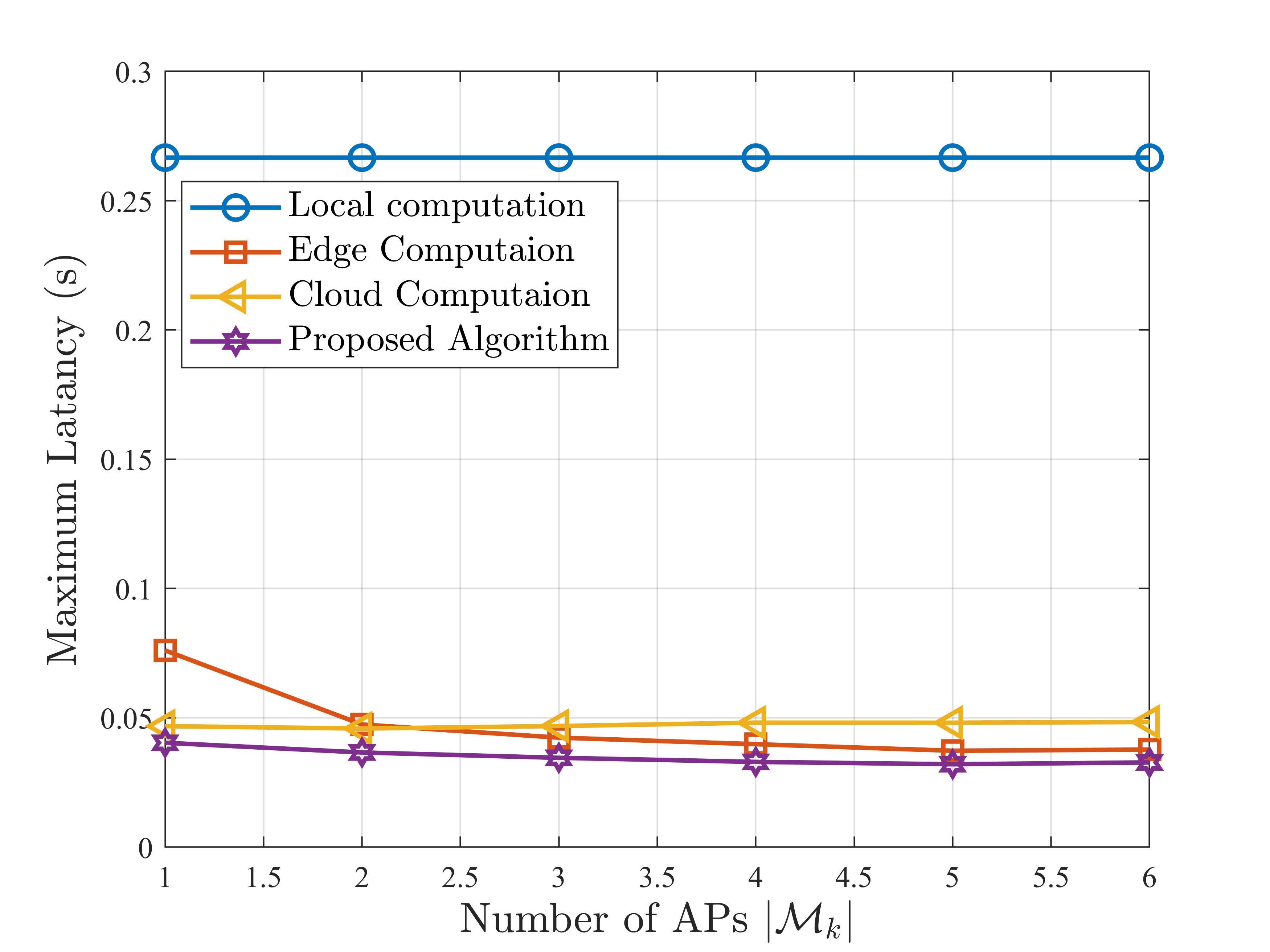}
	\caption{Averaged maximum latency V.S. Cardinality of set \(\mathcal{M}_k\).}
	\label{LatencyVsNum}
\end{figure}

To obtain statistically accurate results, the Monte-Carlo simulation results are obtained by averaging over 100 trials. Fig. \ref{LatencyVsNum} depicts the averaged maximum latency with different number of APs that provide service for each terminal. As excepted, our proposed method is superior over all benchmarks. This is because that the proposed method can flexibly schedule the resources of terminals, mobile edge servers, and cloud sever to obtain the cooperative gain, thereby significantly improving the system performance. Furthermore, the computation latency for MEC, CC, and our proposed algorithm decrease with the increasing number of APs that serve the terminals.This is attributed to the implementation of task partitioning, thereby highlighting the superior efficacy of distributed offloading strategies. However, it is worth noting that the computation latency is slightly increased when the number of APs is larger than 5. This is can be explained by the fact that the significant interference would cause performance degradation on communication rate when more APs are involving in offloading resource, which results in longer computation latency. Therefore, it is beneficial to strike a good balance between communication and computation performance. 

\subsection{Effect of Execution Frequency}

\begin{figure}
	\centering
	\includegraphics[width=3.2in]{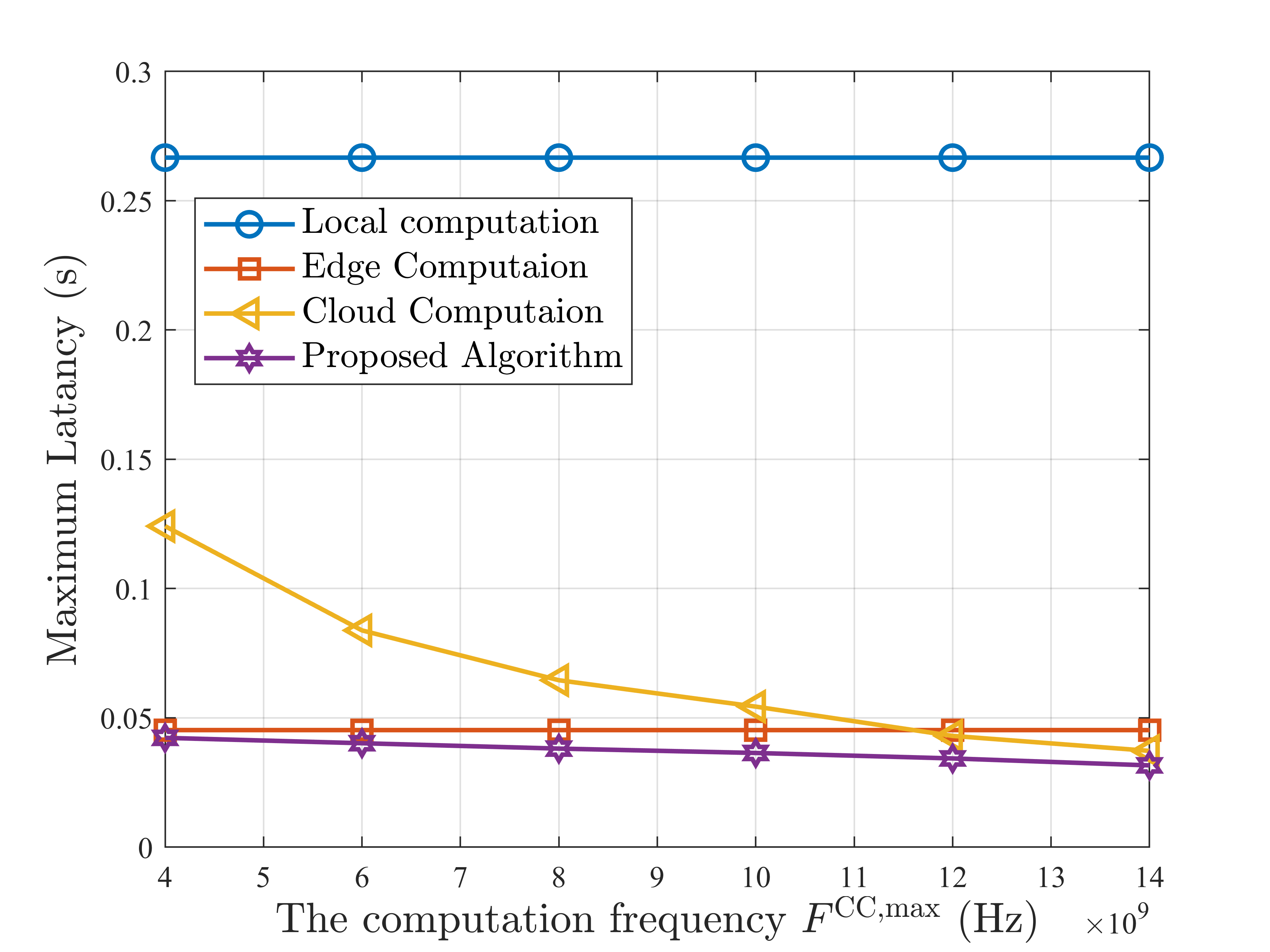}
	\caption{Averaged maximum latency V.S. Execution frequency of cloud server \(F^{\text{CC},\max}\) with \(|\mathcal{M}_k| = 3\).}
	\label{LatencyVsFreq}
\end{figure}
Fig. \ref{LatencyVsFreq} illustrates the averaged maximum latency under various execution frequency of cloud sever \(F^{\text{CC},\max}\). It is observed that the computation latency is decreasing with the increasing execution frequency \(F^{\text{CC},\max}\), as more tasks can be offloaded to cloud sever for shorter computation latency. Furthermore, when the frequency \(F^{\text{CC},\max}\) is less than 10 GHz, it is more beneficial to execute the tasks via mobile edge severs.

\subsection{Effect of Fronthaul Link}

\begin{figure}
	\centering
	\includegraphics[width=3.2in]{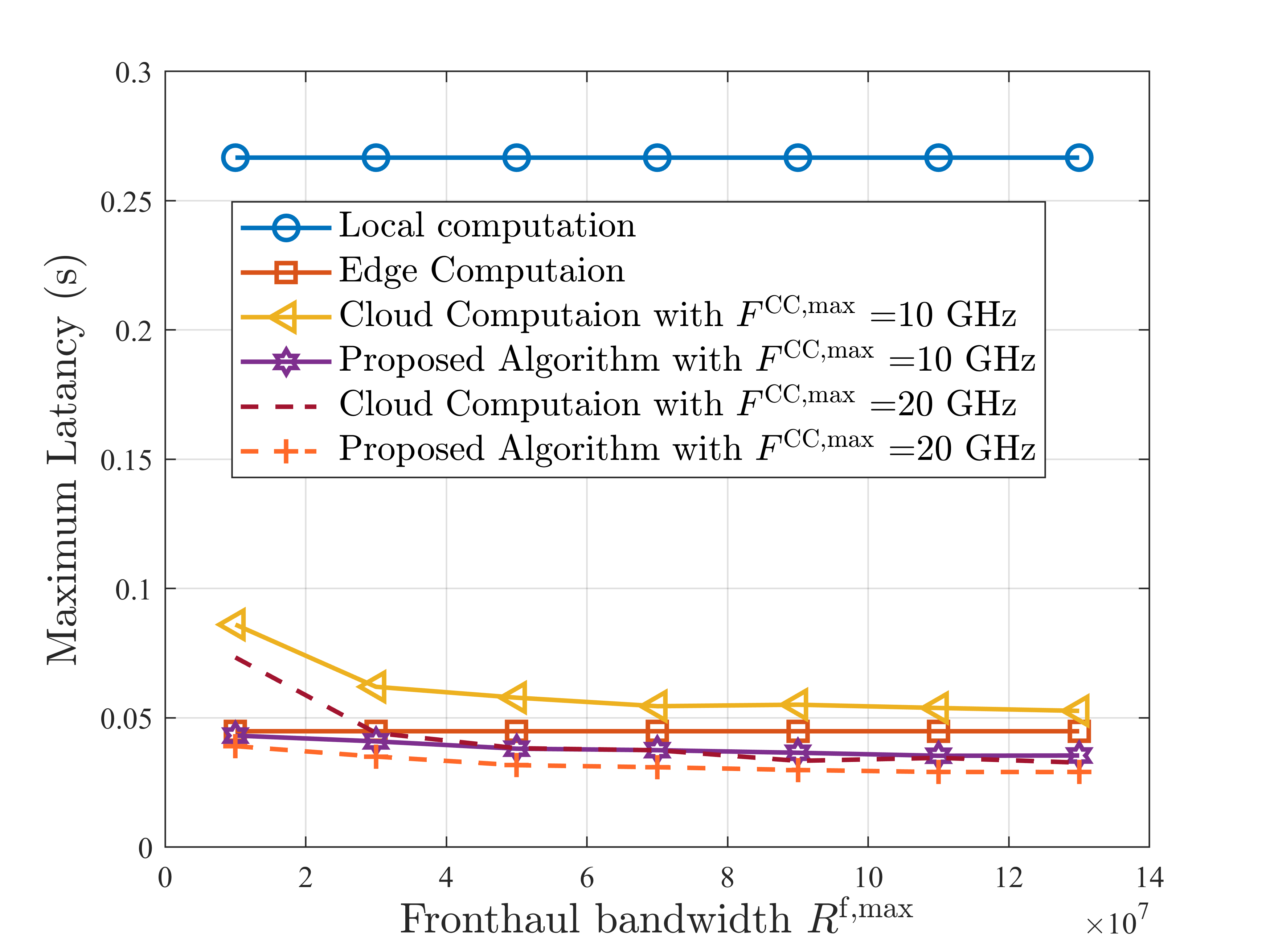}
	\caption{Averaged maximum latency V.S. Fronthaul bandwidth \(R^{f,\max}_m\) with \(|\mathcal{M}_k| = 3\).}
	\label{LatencyVsRf}
\end{figure}

To investigated the impacts of fronthaul bandwidth on the computation latency, Fig. \ref{LatencyVsRf} depicts the averaged latency with various fronthaul capacity over 100 trails. As expected, cloud computation latency decreases notably as the available bandwidth between each AP and the CPU increases. This is due to the fact that a large part of this latency comes from transmission delays, particularly when the fronthaul link has  limited bandwidth. Furthermore, we observe that more tasks are offloaded to the cloud server once the fronthaul link becomes ideal or effectively unbounded, especially for high execution frequency \(F^{\text{CC},\max}\).

\subsection{Effect of Sensing Accuracy}

\begin{figure}
	\centering
	\includegraphics[width=3.2in]{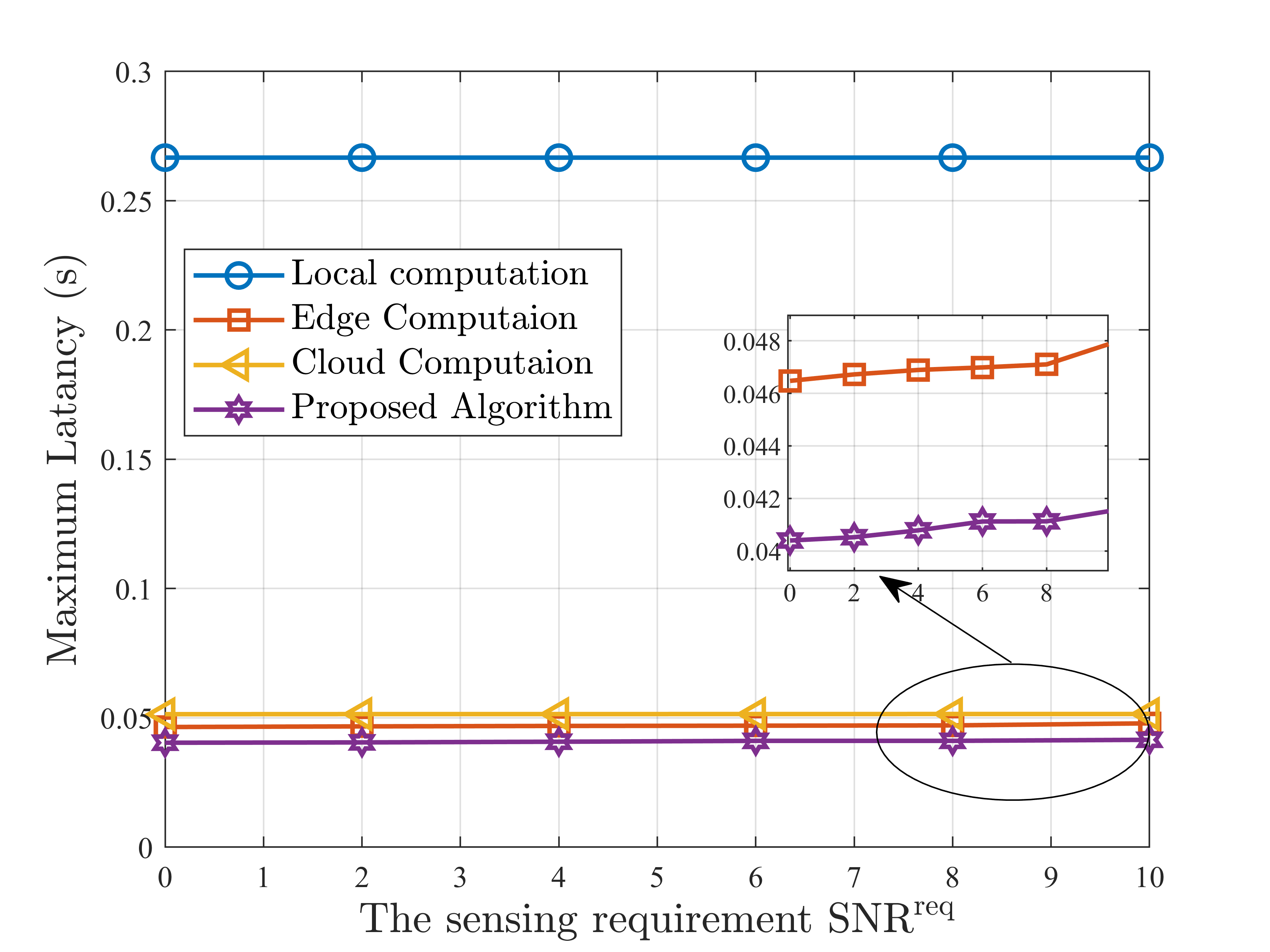}
	\caption{Averaged maximum latency V.S. Sensing requirement \(\text{SNR}^{\text{req}}_k\) with \(|\mathcal{M}_k| = 3\).}
	\label{LatencyVsSINR}
\end{figure}
In Fig. \ref{LatencyVsSINR}, we investigate the impacts of sensing requirements on computation latency by averaging over 100 random generation of terminals' locations. One can observe that the computation latency slight increases with the increasing sensing requirements. This is because that the terminal needs to allocate more power for sensing in order to satisfy  the sensing requirements, and thus less power is left for task offloading. This phenomenon also reflects the trade-off between the sensing and communication functionalities in the ISCC system.

\subsection{Effect of Number of Antennas}

\begin{figure}
	\centering
	\includegraphics[width=3.2in]{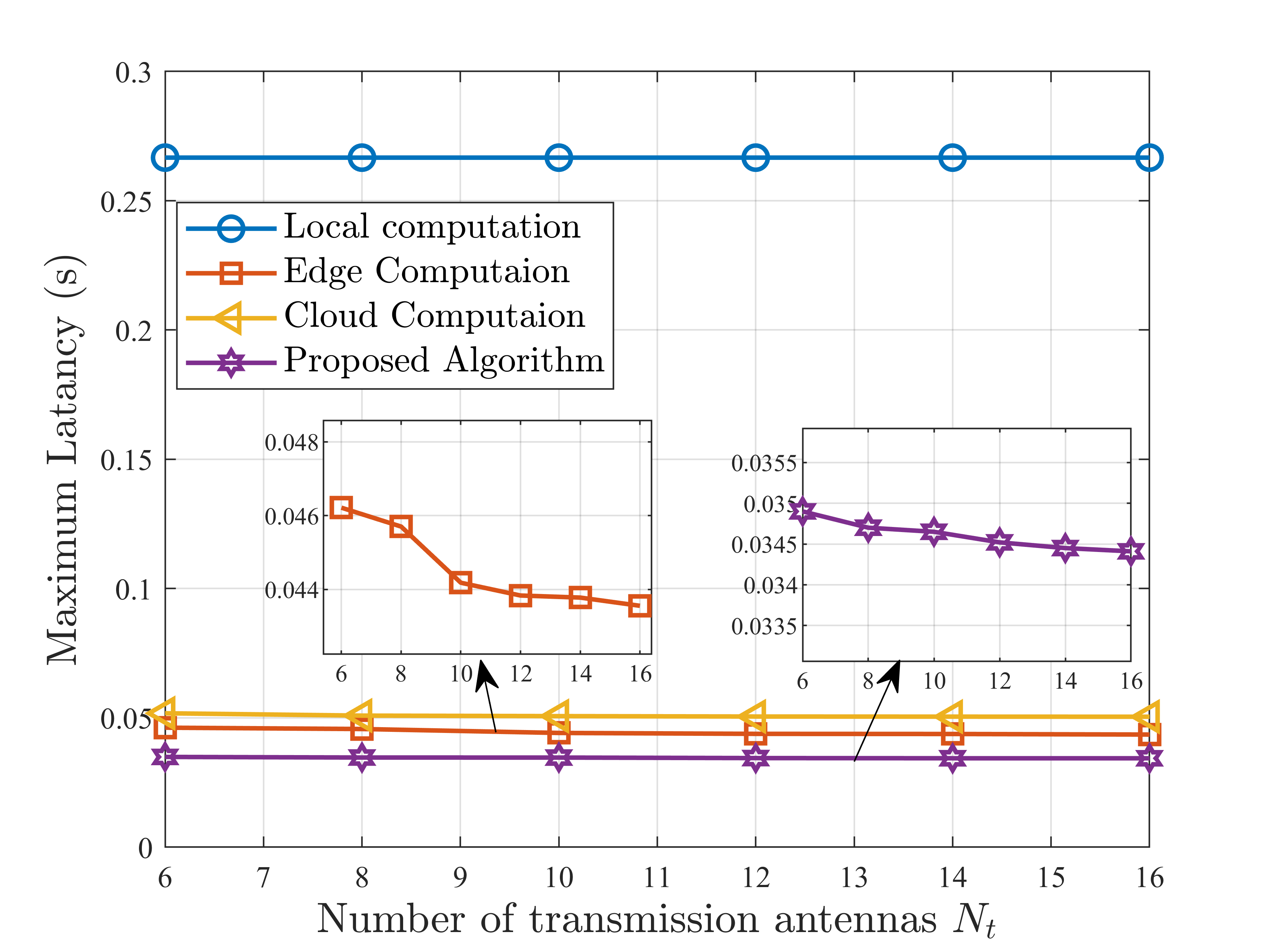}
	\caption{Averaged maximum latency V.S. Number of Antennas \(N_t\) with \(|\mathcal{M}_k| = 3\).}
	\label{LatencyVsN}
\end{figure}

Fig. \ref{LatencyVsN} depicts the averaged latency with different number of transmission antennas \(N_t\). As excepted, the latency decreases when the number of antennas increases, owing to the enhanced transmission or receiving gain. Furthermore, we observe that the performance of our proposed method will approach to that with MEC. This is due to the fact that more tasks can be offloaded to mobile edge servers, rather than cloud sever, with the increasing communication rate.

\section{Conclusion}
In this paper, we investigated a cell-free massive MIMO-enabled ICCS system, where the terminals have the option to offload its local task to either mobile edge servers or cloud server. To characterize the benefits of distributed computation, we formulated a mix-integrate programming and non-convex problem to minimize the maximum latency. To tackle this issue, we decomposed it into three sub-problems. By alternatively devising the offloading decision, ISAC beamforming, execution frequency, and fronthaul bandwidth, our proposed algorithm can rapidly converge to a local optimal solution, which has been verified  by analysis and numerical results. Furthermore, simulation results demonstrated that the trade-off between communication performance and computation latency, as well as the superiority of our proposed method over other benchmarks, verifying its effectiveness and viability for ICCS applications.

\begin{appendices}
\section{Proof of Lemma \ref{upperlemma}}
\label{Prooflemma1}
With the fixed \(\mathbf{v}_{k,m}\) and \(\mathbf{w}_{k,m}\), the optimal \(V_{k,m}\) can be obtained by first-order partial derivative, which is given by 
\begin{equation}
    \label{firstderivative}
    \frac{\partial R_{k,m}(V_{k,m})}{\partial V_{k,m}} = (V^{-1}_{k,m})^T - \text{MSE}^T_{k,m}.
\end{equation}
Then, according to the optimal first-order optimality condition, letting \(\frac{\partial R_{k,m}(V_{k,m})}{\partial V_{k,m}}\) equal to 0, we obtain the optimal \(V_{k,m} = \text{MSE}^{-1}_{k,m}\).

Finally, we complete this  proof by substituting the optimal \(V^{\text{opt}}_{k,m}\) into (\ref{inequalitylem}), and have
\begin{equation}
    \label{proof2}
    \begin{split}
        R_{k,m} &= B\log_2(\text{MSE}^{-1}_{k,m}) \\
        & = B \log_2\Big(\big( 1 - \mathbf{w}^H_{k,m}\mathbf{H}^{H}_{k,m}(\mathbf{N}_{k,m} + \mathbf{H}_{k,m}\mathbf{w}_{k,m}\mathbf{w}^H_{k,m}\mathbf{H}_{k,m})^{-1}\mathbf{H}_{k,m}\mathbf{w}_{k,m}\big)^{-1}\Big) \\
        & \overset{a}{=} B \log_2 (1+\mathbf{w}^H_{k,m}\mathbf{H}^{H}_{k,m}(\mathbf{N}_{k,m})^{-1}\mathbf{H}_{k,m}\mathbf{w}_{k,m}) \\
        & \overset{b}{=} B \log_2 (\mathbf{I}_N+\mathbf{H}_{k,m}\mathbf{w}_{k,m}\mathbf{w}^H_{k,m}\mathbf{H}^{H}_{k,m}\mathbf{N}_{k,m}^{-1}),
    \end{split}
\end{equation}
where \(a\) is obtained via the Woodbury matrix identity and \(b\) relies on the property \(\det(\mathbf{I}+\mathbf{AB}) = \det(\mathbf{I}+\mathbf{BA})\), respectively. 
\end{appendices}
%\subsection{Effect of Blocklength}
%We investigate the effect of blocklength on the weighted sum rate, as illustrated in Fig. . 
%\bibColoredItems{black}{zhi2022power,wu2020joint,guo2022uplink,loyka2001channel}

\bibliographystyle{IEEEtran}
\bibliography{myref}

\end{document}